\documentclass[9pt]{article}
\usepackage[a4paper, total={7.2in, 10in}]{geometry}
\usepackage{caption,subcaption}
\usepackage{url}
\usepackage{float}
\usepackage{algorithm}
\usepackage{algpseudocode}
\usepackage{graphicx}

\usepackage[utf8]{inputenc}
\usepackage[T1]{fontenc}
\usepackage{lineno}

\usepackage[mathscr]{euscript}
\usepackage{bigstrut}
\usepackage{multirow}

\usepackage{amssymb,amsmath,amsthm,amsfonts}
\newtheorem{observation}{Observation}

\usepackage{xcolor}

\newcounter{algsubstate}
\renewcommand{\thealgsubstate}{\alph{algsubstate}}
\newenvironment{algsubstates}
  {\setcounter{algsubstate}{0}%
  \renewcommand{\State}{%
     \stepcounter{algsubstate}%
     \Statex {\footnotesize\thealgsubstate:}\space}}
  {}

\newcounter{protocol}
\makeatletter
\newenvironment{protocol}[1][tbhp]{%
  \let\c@algorithm\c@protocol
  \renewcommand{\ALG@name}{Protocol}
  \def\ALG@step%
  {%
  \addtocounter{ALG@line}{1}%
  \addtocounter{ALG@rem}{1}%
  \ifthenelse{\equal{\arabic{ALG@rem}}{\ALG@numberfreq}}%
      {\setcounter{ALG@rem}{0}\alglinenumber{Step \arabic{ALG@line}}}%
      {}%
  }%
  \begin{algorithm}[#1]%
  }{\end{algorithm}
}
\makeatother

\title{Cascading Failures in Power Grids}

\author{Rounak Meyur}

\begin{document}

\maketitle

Critical infrastructures are defined as \emph{those physical and cyber-based systems that are essential to the minimum operations of the economy and the government} \cite{clinton,gop}. Since they provide crucial support for the delivery of basic services to almost all segments of society, they form the backbone of any nation's economy. As one of the most complex, large-scale networked systems, electric power system has become increasingly automated in the past few decades. However, the increased automation has introduced new vulnerabilities to equipment failures, human errors~\cite{nerc2021,NERC_2016,NERC_2015,NERC_2014,NERC_2013}, weather and other natural disasters~\cite{onl_2019,Weiss2019}, and physical and cyber-attacks~\cite{gop,amin2020}. The ever-increasing system scale and the strong reliance on automatic devices increase the likelihood of turning a local disturbance into a large-scale cascading failure~\cite{liu2021,low2021,Schafer2018,rounak2018,barrett_2012,setola_2008}. This kind of wide-area failure may have a catastrophic impact on the whole society. Reports of recent major power system blackouts~\cite{texas2021,busby2021,victoria2021,canada2022,India,kundur,2003_bout,WSCCout} have shown how several events ranging from minor equipment failure and operator errors to severe weather events (such as forest fires, hurricanes and winter storms) have triggered widespread system wide power disruption affecting millions of customers. This necessitates the development of a framework which would assess the vulnerability of the power grid subjected to any of these events, and thereby allowing energy policy makers to identify critical components in the grid and subsequently allocate budgets to harden them.

Statistical analysis of more than 400 blackouts in USA from 1984 to 1999 indicates that a large blackout, though rare, is more likely to occur than expected (heavy tails of a power law distribution)~\cite{carreras01}. Therefore, large blackouts require more attention not only due to their higher probability of occurrence, but also due to the enormous societal damage caused by such events. Following this observation, several works~\cite{low2021,carreras02,carreras03,carreras04,dobson2,chen_thorpe_dobson,zhang_2016,pahwa,soltan_2014,Bernstein_2012} have proposed multiple failure models to represent the system dynamics leading to a cascading outage. They have studied cascading failures in power grids using quasi-steady state analysis with DC power flow. With any reactive power component being ignored and the assumption of a flat voltage profile, the DC power flow analysis may produce good approximations under some circumstances, e.g., when performing steady-state planning level studies. However, the increased penetration of converter-based generator technologies, loads and transmission devices have contributed to newly evolved dynamic stability behaviors of the power grid~\cite{stability2021}. Major cascading outages are caused when transient rotor angle stability and voltage stability of the power grid are affected~\cite{kundur,hines_2010,hines_2010_other,hines_2016}. Therefore, a simple cascading failure model based on DC power flow analysis is not a suitable tool to simulate such events. In this paper, we consider the AC power flow model to accurately simulate the actual operating point in the power system.

Several physics-based models have been used to study cascading failures in power grid networks and interdependent power and communication networks~\cite{stanley_2010,roni_2010,stanley_2011,raissa_2012,stanley_2018,hill2020,Son_2012,buldyrev2020}. The authors have considered the effect of connectivity between layered networks on the cascade probability in each network, and used the sandpile dynamics~\cite{sandpile2} to represent the cascade tripping of loads in the power grids. These papers are useful in that one can often either obtain analytical results, or carry out large number of simulations to get a detailed understanding of cascade dynamics. The physics based models are simplified models capable of showcasing mechanistic possible behavior of complex network systems, rather than providing precise predictions which requires engineering models with a large number of parameters~\cite{buldyrev2020}. The models fail to replicate the actual system conditions in a power grid where a node (or bus) trips due to under-voltage or under-frequency and not due to overload. Further, stability of a power system subjected to cascading events is evaluated either from the network structure point of view (evaluating the degree distribution of nodes) \cite{raissa_2012,stanley_2010,stanley_2011,stanley_2018} or from the convergence of steady-state power flow solution~ \cite{carreras02,carreras03,carreras04,dobson2,chen_thorpe_dobson,zhang_2016,pahwa}. Such measures do not necessarily cover all possibilities of grid instability~\cite{stability2021}, as non-linear mechanisms such as rotor angle stability or voltage collapse are not accurately captured in these methods \cite{hines_2010}. 
In this work, dynamic transient analysis has been used to assess stability of the power system.
\begin{figure*}[tbhp]
\centering
\includegraphics[width=0.9\textwidth]{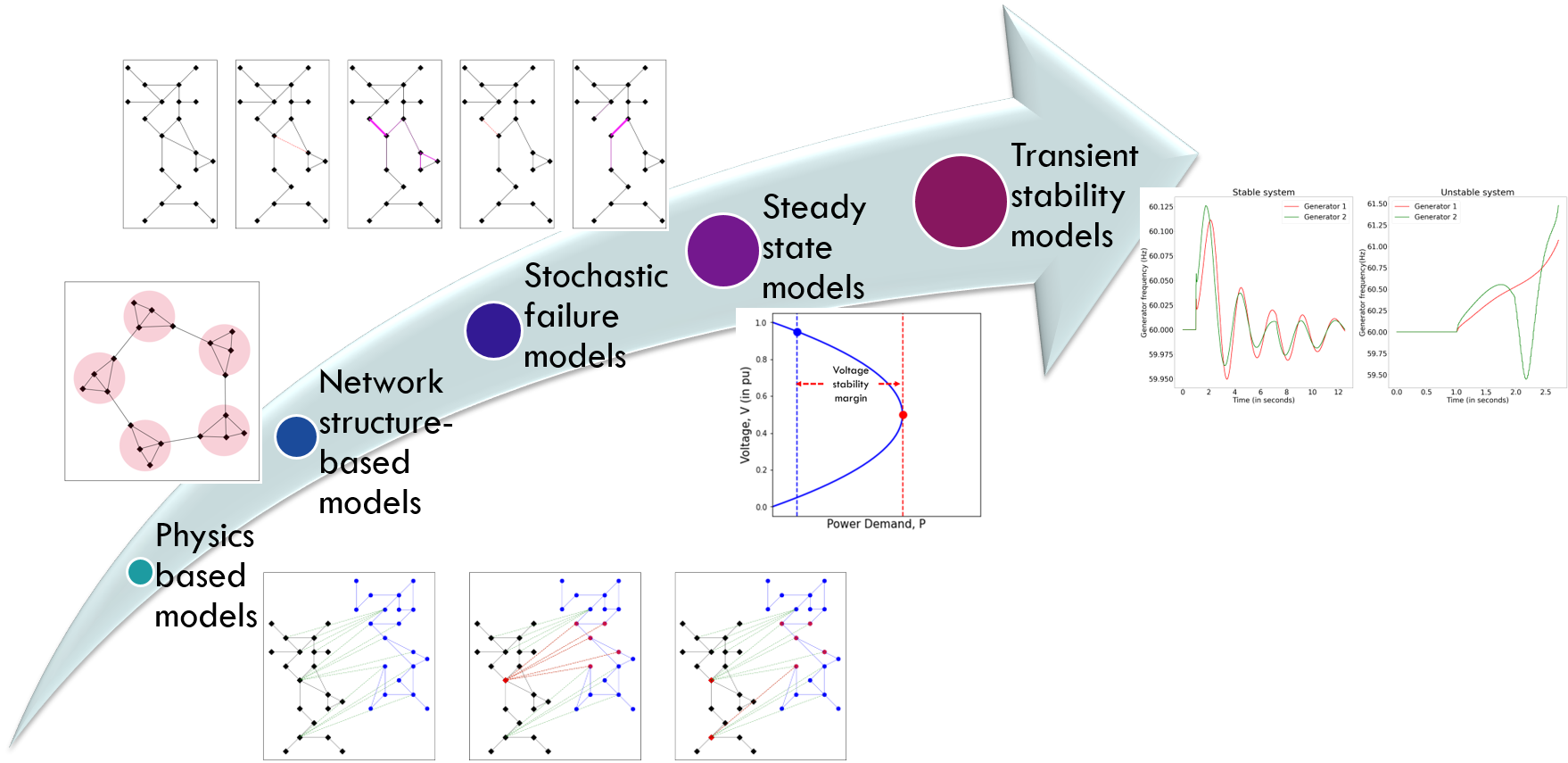}
\caption{Summary of different failure models in literature -- physics based models represent cascades in inter-dependent networks where dependent nodes in either network fail simultaneously, network structure-based failure models proposes importance of network motifs, stochastic failure models assign probabilities to edges adjacent to failed nodes/edges and cascade propagates. The steady-state and time domain simulation models depict a more realistic version of cascading outages. The operation of protection systems is not considered in any of these works.}
\label{fig:lit-models}
\end{figure*}

The reports of certain major blackouts~\cite{2003_bout,kundur} suggest that cascades need not propagate locally due to the complex non-linear nature of the power grid. Furthermore,~\cite{WSCCout} discusses the various reasons leading to the historic 1996 WSCC outage, the most important being the operation of relays.  Based on the NERC data, in more than $70\%$ of the major disturbances, failures in protective relays are found to be a contributing factor~\cite{chen_thorpe_dobson}. Among these failures, a failed protection system that remains dormant in normal operating conditions and becomes exposed when an abnormal condition in the system forms, is the most troublesome to tackle \cite{tamronglak1}. Such failures are termed as hidden failures and these are capable of causing widespread cascading failures in the power system network leading to a major blackout \cite{tamronglak2}. 
This is equivalent to the human immune system where an immune response following immunization might be more damaging than the pathogen it is supposed to protect against~\cite{simon2020}.
It is evident from the above discussion that protection systems play a key role in cascading events. Though most of the papers consider line outages due to overload, the protection system in the power network respond to measured impedance, voltage and current. The vital contribution of the proposed work is the inclusion of a stochastic model to simulate hidden failures in the power system whose effects surface in the aftermath of a human initiated attack on the network.

An important step in modeling cascading failures is to evaluate the probability of tripping of each component in the power system network. The NERC statistics over the past decade~\cite{NERC_2013,NERC_2014,NERC_2015,NERC_2016,nerc2021} show that relay misoperations due to unnecessary trips are more probable than failure to trip. Such relay failures or faulty settings are often the principal determinant of the occurrence of a hidden failure in the power system. In this work the AC power flow model is used to obtain the system conditions at each instant of simulation and transient stability analysis is performed to assess the stability of the grid. The operation of protection systems for generators, transmission lines and transformers is modeled along with the stochastic occurrence of hidden failures in them. The trip signals of these relays are considered as the sole contributors of node and edge outages in the network.

Another important aspect of studying cascading events in the power system network is the impact of different initiating events. For example,~\cite{raissa_2012,stanley_2010,zhang_2016} have initiated cascaded failures by targeting random node(s) in the network. On the contrary, cascaded outages triggered by weather events are initiated by targeting geographically correlated nodes~\cite{Weiss2019}. Given the complex non-linear nature of the power system, the \emph{optimal critical set} problem is worth mentioning~\cite{rounak2018}. The goal is to find the optimal set of nodes that results in maximum damage to the network. The results show that a greedy choice of high voltage (500kV) nodes leads to significant impact on the power network, often resulting in an unstable system causing widespread power outages. 

\noindent\textbf{Contributions.}\ In this work, we build a framework to analyze cascading failures on a given power network that has been subjected to physical attacks on multiple nodes. We compare the extent of cascading outages in case of a detonation of a large tactical device in Washington DC, and a strategic targeted physical attack performed simultaneously on different power system substations located far apart from each other. The following are the main conclusions of our analysis: (a) A time-domain, AC analysis is essential in capturing the full effects of a cascading event on the electric power grid. (b) A strategic targeted attack on few critical substations (as few as $2$) is capable of leading the power system to collapse within a few seconds. 
(c) In context of a strategic attack on a critical target set, the addition of target nodes to an existing target set does not necessarily increase the impact on the power grid. 
(d) The load-generation balance plays a key role in the extent of cascading outages.

\section*{Proposed Framework}
Protocol~\ref{alg:framework} lists the steps required for using our proposed framework. The first step is to obtain the information regarding the power grid network (Step 1). This is stored as a graph network with a list of edges and nodes and their associated parameters. Thereafter, the generator control system is modeled from the parameters of the governor and excitation system (Step 2). The next step (Step 3) is to define the deterministic $1/0$ logic for the protection relays in the power grid. The list of hidden failures are also identified in this step. Then, the initiating event is to be defined (Step 4). Note that our proposed framework can be used for analyzing impact of cascading failures for any given contingency. To this end, we need to provide the framework with the list of nodes which are directly impacted by the event. We model the event as a three phase fault on these nodes to initialize the cascading failure analysis. Finally, we use Algorithm~\ref{alg:ac-cascade} to compute the impact of the initiating event on the power grid (Step 5).
\begin{protocol}
	\caption{Outline of the proposed framework}
	\label{alg:framework}
	\begin{algorithmic}[1]
		\State Construct the power grid network.
		\begin{algsubstates}
		\State Get the list of labeled edges from the table of transformers and transmission lines.
		\State Get the list of labeled nodes from table of substations, generators and load buses.
		\end{algsubstates}
		\State Model the generator control systems.
		\begin{algsubstates}
		\State Get parameters of generator governors.
		\State Get parameters of generator excitation systems.
		\end{algsubstates}
		\State Model the protection relays.
		\begin{algsubstates}
		\State Define protection logic for generators, transmission lines and transformers.
		\State Identify list of relays with one or more hidden failures.
		\end{algsubstates}
		\State Model the initiating event.
		\begin{algsubstates}
		\State Evaluate the list of nodes affected by the event.
		\State Model the initiating event as fault in power grid.
		\end{algsubstates}
		\State Evaluate impact of the event.
		\begin{algsubstates}
		\State Perform transient stability analysis.
		\State Evaluate operation of protection relays.
		\State Compute the total number of node outages.
		\State Determine the stability of power grid.
		\end{algsubstates}
	\end{algorithmic}
\end{protocol}

\subsection*{AC Power Flow Analysis}
The power grid under consideration is represented as a fully connected undirected graph $\mathscr{G}(\mathscr{V},\mathscr{E})$. $\mathscr{V}$ and $\mathcal{E}$ respectively denote the sets of nodes (substations/buses) and edges (transmission lines and transformers) in the network. The power flowing through each edge $e\in\mathcal{E}$ is constrained by a pre-determined thermal limit of the line. The complex voltages at each node $v\in\mathcal{V}$ in the network represent the states of the power system. The goal of the AC power flow analysis is to compute the voltage magnitudes and phase angles at each node using the measured active and reactive power flows and injections~\cite{pskundur}.

\subsection*{Transient Stability Analysis}
The ability of a power system to maintain synchronism when subjected to a large disturbance is termed as transient stability~\cite{pskundur,sauer,stability2021}. In response to a rapid loss of load (or generation), the power system frequency will increase (or decrease).  However, the generator controls respond to this change by changing the power output to meet the electric load demand based on a set of differential equations. In the present study we have followed a numerical integration method to solve these equations. This means that for each time instant the AC power flow problem is solved to obtain the states of the power system. This solution is used as initial values for the differential equations required to solve the transient stability problem.

\subsection*{Power System Collapse}
A power system collapse can occur due to different reasons that can be attributed to the transient rotor angle stability and voltage stability of the grid. In the present study, the transient rotor angle stability is used to identify a system collapse. If a set of generator rotor angles differ from the rotor angles of another set by more than $180$ degrees, the two sets of generators are said to be operating out of step~\cite{Phadke-Horowitz}. In such a scenario, the generators trip due to the operation of out-of-step relays causing load-generation imbalance in the power grid. This causes frequency to drop below the allowable range and thereby triggers automatic under-frequency load shedding. Since this load-shedding is automatic, it can result in further load-generation imbalance. For example, over-frequency at the less loaded generator buses can cause the generators to trip and this process, if allowed to continue, can result in a widespread blackout.

\subsection*{Operation of Protection Systems}
The protection system in a power system detects faults and issues a trip signal to separate the faulted section from the healthy section. These protective relays play an important role in cascading outages since these are solely responsible for tripping edges and nodes in the network. We consider a generic protection system $\mathsf{P}_i$ that protects an element $i$ (node or edge) and operates based on measured quantity $W_{\mathsf{P}_i}$. Examples of measured quantities are current, voltage and impedance. Each protection system has a zone of operation, denoted by $\mathscr{R}_{\mathsf{P}_i}$. The trip decision $u_{\mathsf{P}_i}=\{0,1\}$ issued by the relay is a binary quantity where $u_{\mathsf{P}_i}=1$ represents that the protected element $i$ (node or edge) is disconnected and vice-versa. The element $i$ is disconnected or \emph{tripped} if the measured quantity $W_{\mathsf{P}_i}$ encroaches the zone of operation. That is, $u_{\mathsf{P}_i}=1$ if $W_{\mathsf{P}_i}\in\mathscr{R}_{\mathsf{P}_i}$, and $u_{\mathsf{P}_i}=0$ otherwise

The zone of relay operation is contingent upon a set of conditions occurring simultaneously. Such conditions are physically realized through electro-mechanical/digital contacts, such that the relay issues a trip signal when all of them are satisfied. When a relay has a hidden failure in one or more of these contacts, only certain conditions are required to be satisfied to send the trip signal. Under such circumstances, its zone of operation alters to $\mathscr{H}_{\mathsf{P}_i}$ where $\mathscr{R}_{\mathsf{P}_i}\subset\mathscr{H}_{\mathsf{P}_i}$. For example, a directional overcurrent relay issues a trip signal to the circuit breaker if an overcurrent is detected in a particular direction. This is designed through two contacts - overcurrent and directional contacts. If a hidden failure occurs in the directional contact, the relay issues a trip signal as soon as it detects an overcurrent irrespective of the direction. Further, such a failure remains hidden because the relay does not issue any trip signal until an overcurrent is detected. Other details regarding hidden failures and their way of occurrence in different relays are provided in the SI.

Drawing an analogy to the human immune response, the action of protective relays in a power grid is similar to the role played by active immunity developed in a human through direct clinical infection or by specific immunization. The antigens of infected cells are detected either by B-cells (humoral immunity) to create antibodies or by T-cells (cellular immunity) to initiate a chain of response required to neutralize the microbe or its toxin. In a similar fashion, a protective relay detects a fault in a power system and sends a trip signal to the circuit breaker in order to isolate the faulted section. An unwanted trip caused by a hidden failure in a protection relay is equivalent to an immune system which attacks an innocuous substance or its own uninfected cells causing \emph{anaphylaxis}~\cite{simon2020,park}.

The hidden failures in relays have been modeled using a stochastic approach, wherein a set of relays with hidden failure are randomly sampled from a probability distribution. Let $\mathscr{K}=\left\{\mathsf{P}_{h_1},\mathsf{P}_{h_2},\cdots,\mathsf{P}_{h_n}\right\}$ denote the set of $n$ relays with hidden failures that are randomly sampled from the entire set of relays.

\subsection*{Cascading Failure Model}
The cascading failure model uses an AC power flow based time domain analysis to evaluate the states of the power system at each time instant. The simulation is initiated at time $t=0$ with initial number of nodes $N_0=|\mathscr{V}|$ (where $|\cdot|$ denotes cardinality of set) and condition of power system $c=0$ (denoting stable system). 
At $t=t_f$, the target set $\mathscr{S}$ is attacked, which is done by simulating a three phase fault at the targeted nodes. The connected edges experience an outage due to the targeted attack. Thereafter, the time domain simulation is carried on until $t=t_{\textrm{end}}$ seconds with a time step of $\Delta t$. At each time step, the operation of protective relays is monitored to identify tripped nodes and edges. Additionally, any isolated nodes in the network are considered as node outage. The simulation is stopped if a power system collapse occurs in between or if time reaches $t=t_{\textrm{end}}$ and the condition of power system is altered to $c=1$. Algorithm~\ref{alg:ac-cascade} depicts the pseudocode for the proposed AC transient analysis based cascading failure model. 

\begin{algorithm}
	\caption{Cascading failure model}
	\label{alg:ac-cascade}
	\textbf{Input} network topology ($\mathscr{G}\left(\mathscr{V},\mathscr{E}\right)$), network parameters and settings, relays with hidden failures ($\mathscr{K}$), target set ($\mathscr{S}$), time of attack ($t_f$), time increment ($\Delta t$), simulation end time ($t_{\textrm{end}}$).
	\begin{algorithmic}[1]
		\State Initialize time instant $t=0$.
		\State Initialize condition of power system $c \leftarrow 0$.
		\State Initialize number of nodes $N_0 \leftarrow |\mathscr{V}|$.
		\State Create three phase fault at $t=t_f$ on all nodes in $\mathscr{S}$.
		\While {time instant $t\leq t_{\textrm{end}}$}
		\State Perform transient stability analysis.
		\State Identify tripped edges $\mathsf{F}(t)=\{e\in\mathcal{E}\big|u_{\mathsf{P}_e}=1,\forall \mathsf{P}_e\}$.
		\State Remove tripped edges from graph $\mathcal{E}\leftarrow\mathcal{E}\setminus\mathsf{F}(t)$.
		\State Identify tripped nodes $\mathsf{G}(t)=\{g\in\mathcal{V}\big|u_{\mathsf{P}_g}=1,\forall \mathsf{P}_g\}$.
		\State Remove tripped nodes from graph $\mathcal{V}\leftarrow\mathcal{V}\setminus\mathsf{G}(t)$.
		\State Identify isolated nodes $\mathsf{H}(t)=\{v\in\mathcal{V}\big|\mathsf{degree}(v)=0\}$.
		\State Remove isolated nodes from graph $\mathcal{V}\leftarrow\mathcal{V}\setminus\mathsf{H}(t)$.
		\If {power system collapse occurs}
		\State Change condition to unstable $c \leftarrow 1$.
		\State Compute node outages $\Delta N \leftarrow N_0 - |\mathscr{V}|$.
		\State Stop the process.
		\EndIf
		\State Increment time $t\leftarrow t+\Delta t$
		\EndWhile
		\State Compute node outages $\Delta N \leftarrow N_0 - |\mathscr{V}|$
	\end{algorithmic}
	\textbf{Output} number of node outages ($\Delta N$), condition of power system ($c$).
\end{algorithm}

\section*{Attack Scenarios}
\begin{figure*}[tbhp]
    \centering
	\includegraphics[width=0.48\textwidth]{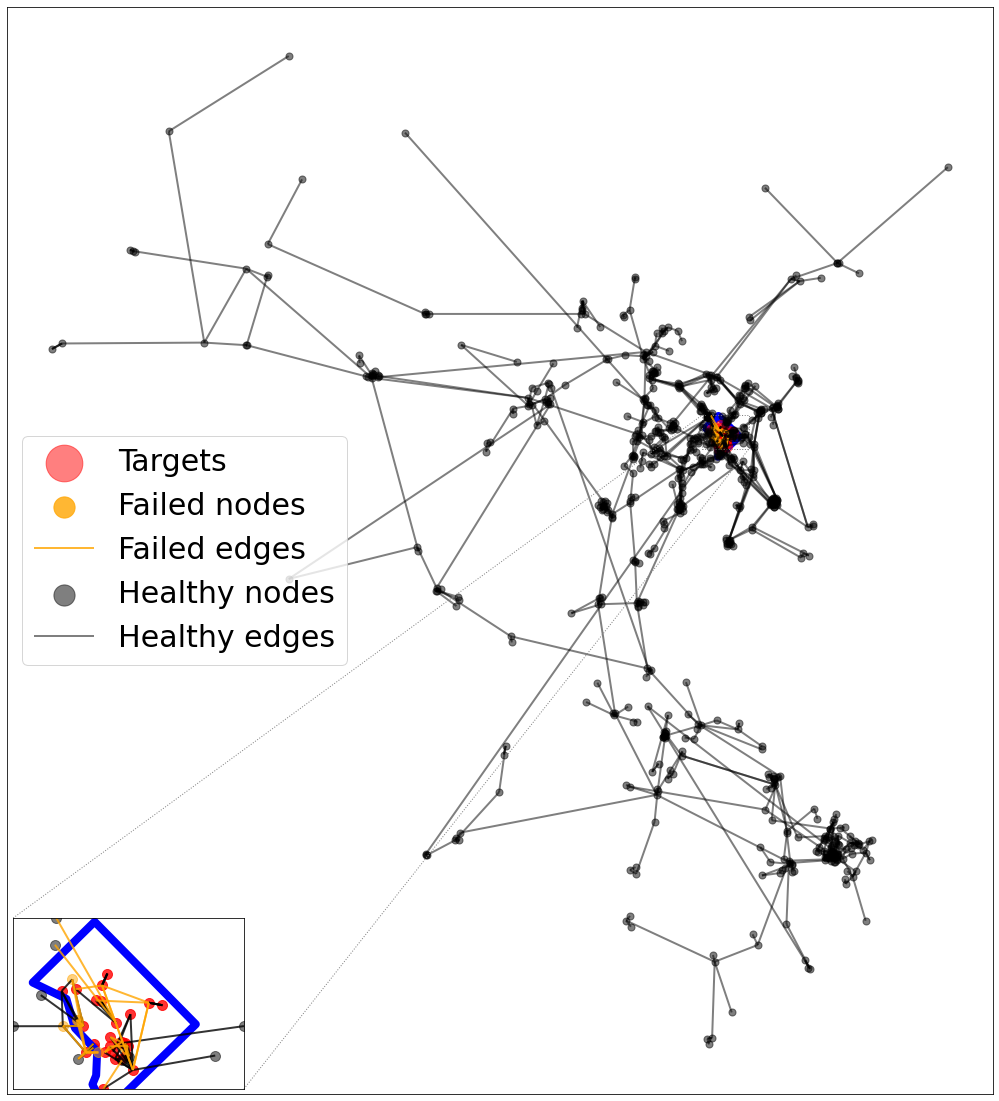}
	\includegraphics[width=0.48\textwidth]{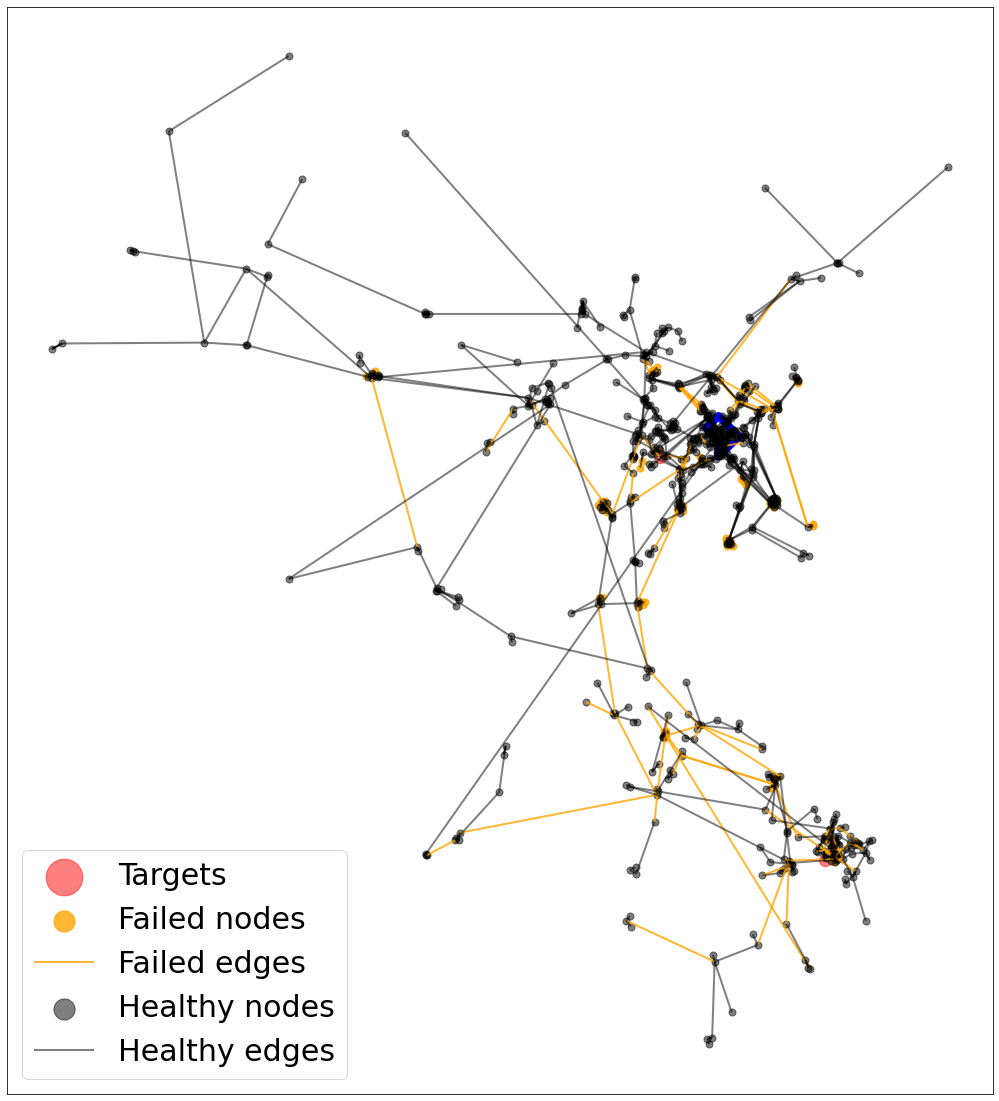}
	\caption{Figure showing a large scale physical attack on Washington DC (left) and targeted attack on strategically selected substation nodes (right). Blue boundary shows Washington DC, red nodes are physically damaged due to the attack and cascading failure propagates to the adjacent orange nodes and edges which trip due to operation of protective relays.}
	\label{fig:geogattack}
\end{figure*}

The targeted adversarial attack considered here is a detonation of a bomb at one or more substations located in and around Washington DC, USA. This attacks depicts a scenario that is aimed at harming the power grid and thereby \emph{indirectly} affecting the human populace of the city. Since multiple substations can be targeted at the same time, this attack scenario can also be called a coordinated targeted attack. We present two different types of targeted attack on the power grid of Washington DC and its neighboring areas.
\begin{itemize}
    \item \emph{Type 1}: a large scale attack caused by detonation of a tactical device which results immense infrastructural damage in the attacked region, and 
    \item \emph{Type 2}: a simultaneous strategic targeted attack on critical substations with an aim of creating a cascading failure throughout the power network.
\end{itemize}
The first scenario can be considered as a large bomb blast in downtown Washington DC. This results in an immense loss of property. However, it needs to be examined whether the disturbance created in the power grid following the event results in a cascading outage in the neighboring regions. The second scenario is a more planned attack on selected substations with the aim of maximizing impact by creating a large scale power outage throughout the grid. The property damage may be limited within the boundaries of the targeted substations; however, the cascaded outage, if one occurs, will severely affect the societal infrastructures in and around Washington DC. There can be multiple possible choices of targeted attack and therefore, we had studied the \emph{optimal critical node} problem in~\cite{rounak2018}. We have summarized the result in the Appendix. We had observed that a greedy choice of high voltage substations (500kV) leads to a significant impact on the power grid. Fig.~\ref{fig:geogattack} shows the two scenarios and the resulting impact obtained after analyzing each of them using the cascade failure model. We notice that for an attack of \emph{Type 1}, the cascaded outages are contained within the boundaries of Washington DC. However, a \emph{Type 2} targeted attack on a single critical node outside Washington DC has resulted in widespread cascading outages.

We analyze each of these attack scenarios through the proposed AC transient analysis based cascading failure model (Algorithm~\ref{alg:ac-cascade}). At the beginning of each scenario, we randomly identify a set of relays with hidden failures denoted by $\mathscr{K}$. This is the only stochastic aspect of the simulation, following which the cascading failure model proceeds in a deterministic fashion. In order to cater for the stochastic presence of hidden failure, we run each scenario multiple times (20 times in our case). We model each attack as a three phase fault on the set of targeted nodes. The protection relays operate based on their settings and causes the following outages.

\section*{Results and Discussion}
\begin{figure}[ht]
	\centering
	\includegraphics[width=0.9\textwidth]{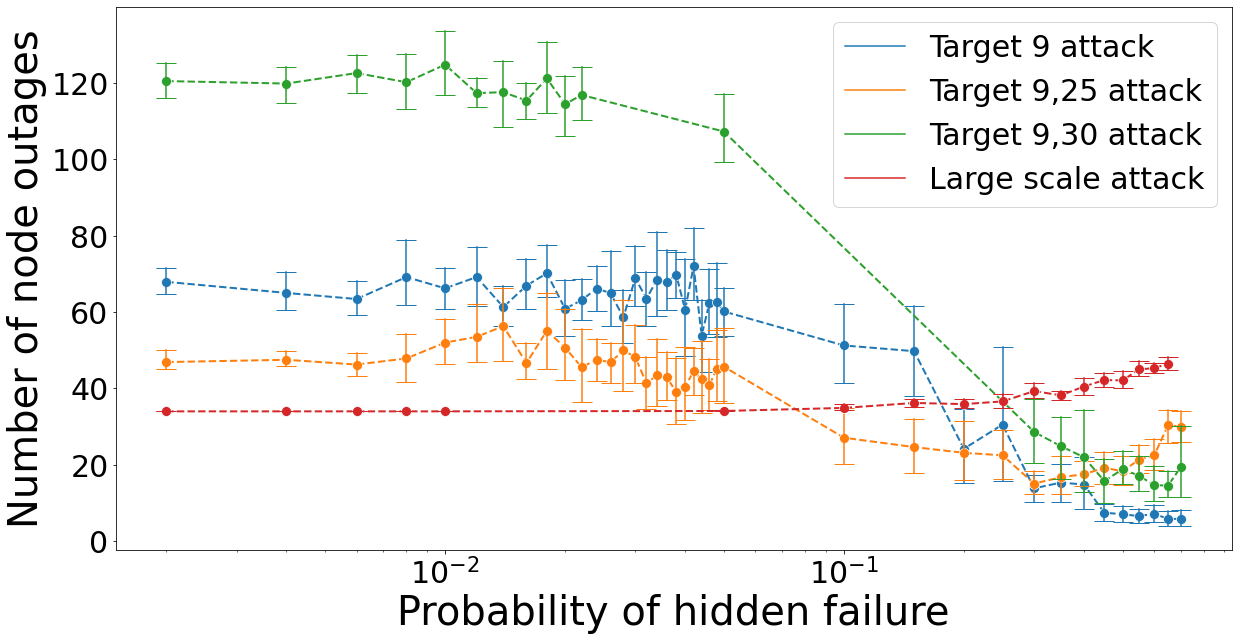}
	\caption{Plot showing variation sensitivity of number of cascading node failures to the probability of hidden failures in protection relays. The node outages caused due to a \emph{Type 1} attack on Washington DC is almost immune to the occurrence of hidden failures. The cascading node outages caused by a targeted attack reduces with increase in hidden failure probability.}
	\label{fig:sense}
\end{figure}

Fig.~\ref{fig:sense} shows the number of node outages which follows a large scale \emph{Type 1} attack and three different \emph{Type 2} attacks on different critical sets of 500kV substations. We list down the key observations here.
\begin{enumerate}
    \item A \emph{Type 1} attack on Washington DC does not result in a cascading power outage outside the boundary of DC. 
    \item Increasing the number of nodes in the target set for a \emph{Type 2} attack does not increase the resulting impact of attack.
    \item An increase in hidden failure occurrence probability reduces the extent of cascading outages for \emph{Type 2} attack on the Washington DC grid.
\end{enumerate}

\begin{observation}
A \emph{Type 1} attack on Washington DC is less likely to cause cascading outages as compared to a strategic targeted attack of \emph{Type 2} on selected 500kV substations.
\end{observation}
We notice that a \emph{Type 1} attack has the minimum impact in terms of the number of node outages in the power grid. However, for a targeted attack of \emph{Type 2}, the number of node outages is significantly higher and often results in system collapse. Washington DC is an area with high load consumption due to large number of residential and commercial establishments; but it does not have enough generating capacity within its geographic boundary. Therefore, there is a significant amount of power which is imported from the neighboring regions (see SI for detailed power flow values).

\begin{figure}[tbhp]
    \centering
	\includegraphics[width=0.46\textwidth]{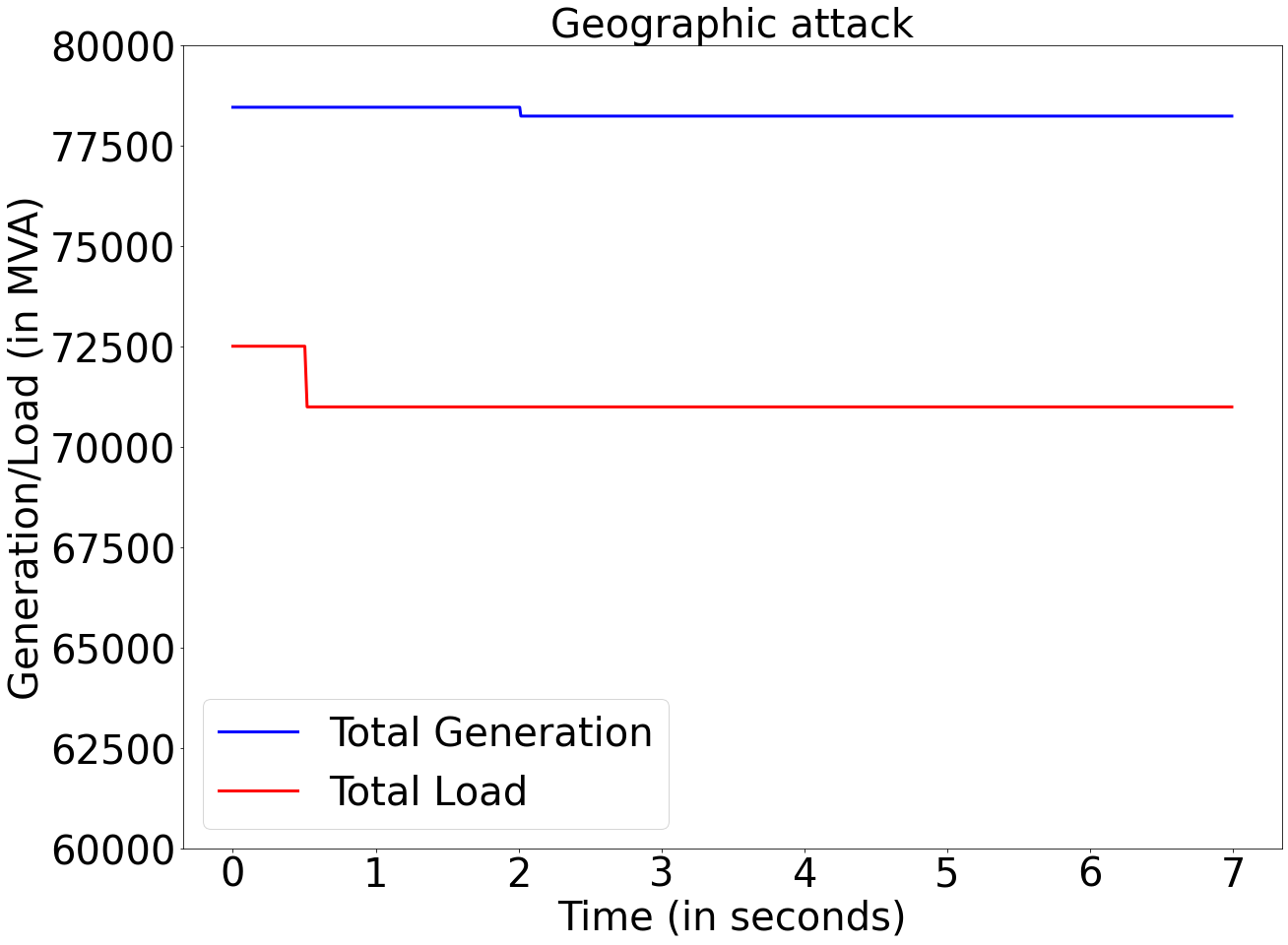}
	\includegraphics[width=0.46\textwidth]{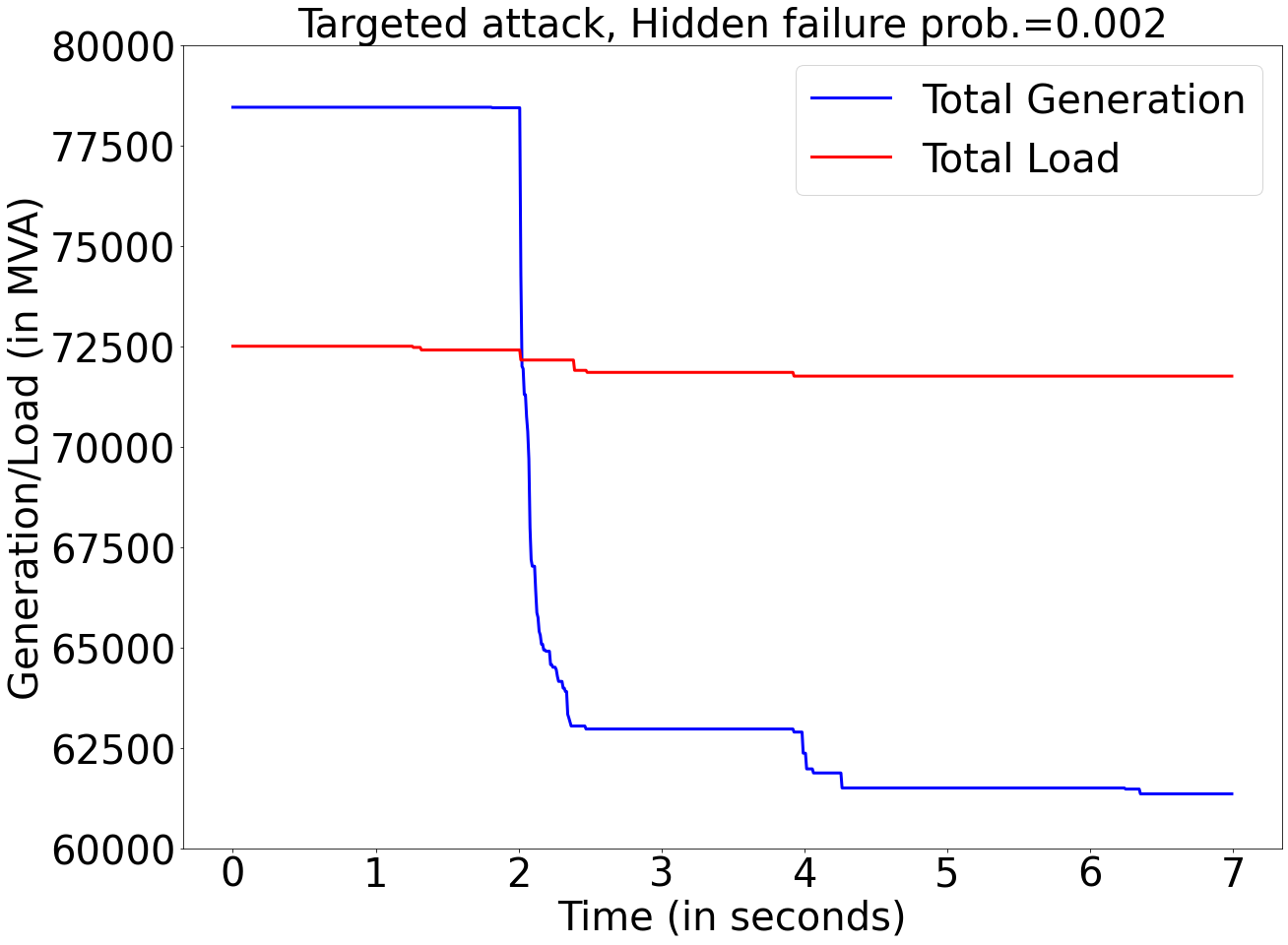}
	\caption{Plots showing change in total system load and generation in the aftermath of a \emph{Type 1} attack on Washington DC (left) and a targeted attack of \emph{Type 2} on a strategically selected substation node (right). For the \emph{Type 1} attack, there is a large drop of load (nearly 2000 MW). For the \emph{Type 2} attack, there is no immediate load drop; rather we observe massive loss of generation due to instability.}
	\label{fig:geog-tar-comp}
\end{figure}

Fig.~\ref{fig:geog-tar-comp} compares the loss of generation and load in the power grid for the two types of adversarial attack. In case of a \emph{Type 1} attack in Washington DC, there is a large loss of load within the boundary of the city. Therefore, little to no power is required to be imported from the neighboring regions. On the contrary, a targeted attack on selected substations along one of the power import paths causes outage of \emph{important} transmission lines connecting Washington DC to the neighboring regions, thereby interrupting the power flow along them. The outage of significant 500kV lines result in a reduced power support in the power system. The deficit needs to be supplied by the generating units along the other import paths. This results in transient rotor angle and voltage instabilities in the power grid and further loss of generation.

\begin{figure}[tbhp]
    \centering
	\includegraphics[width=0.46\textwidth]{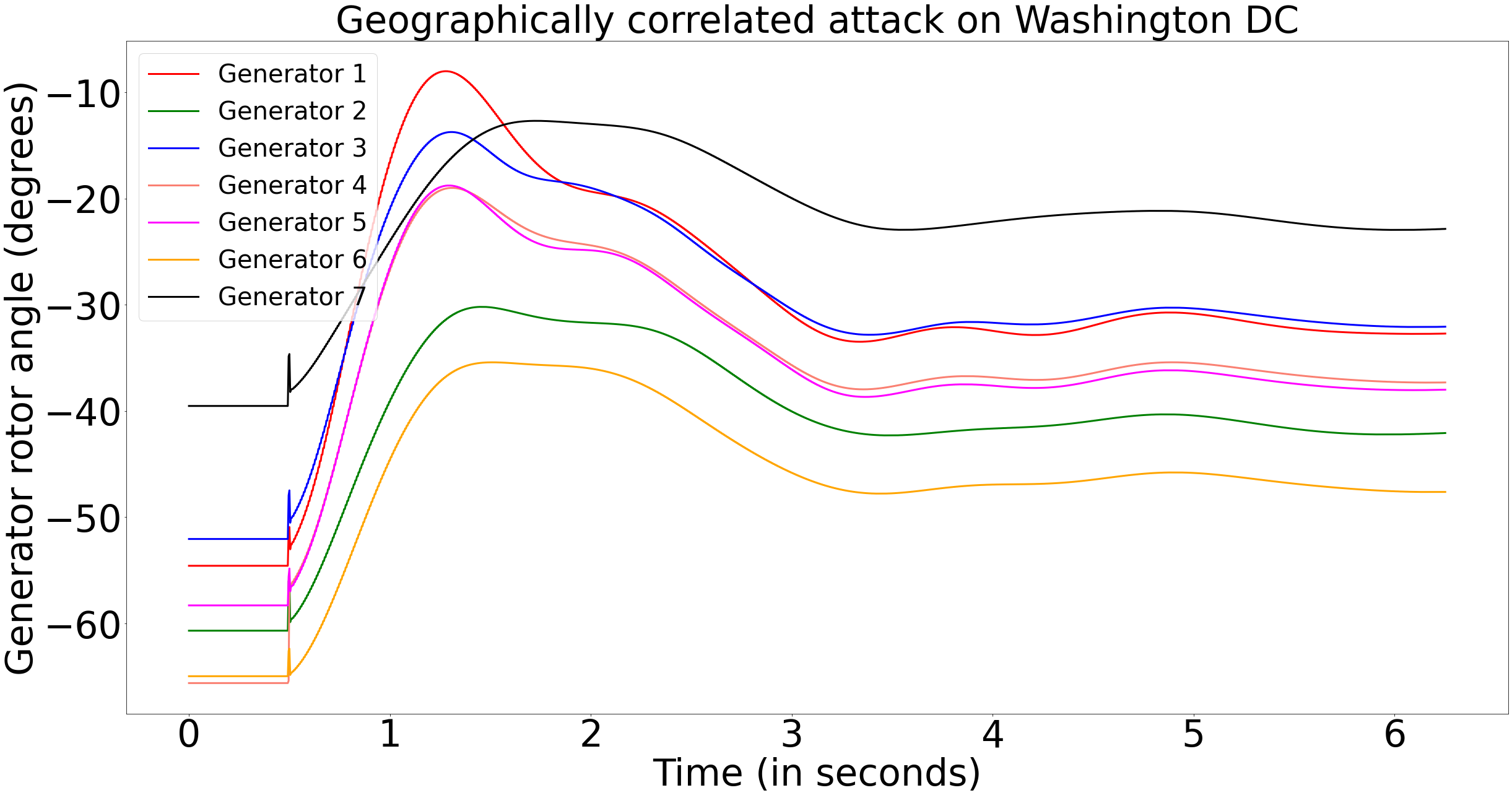}
	\includegraphics[width=0.46\textwidth]{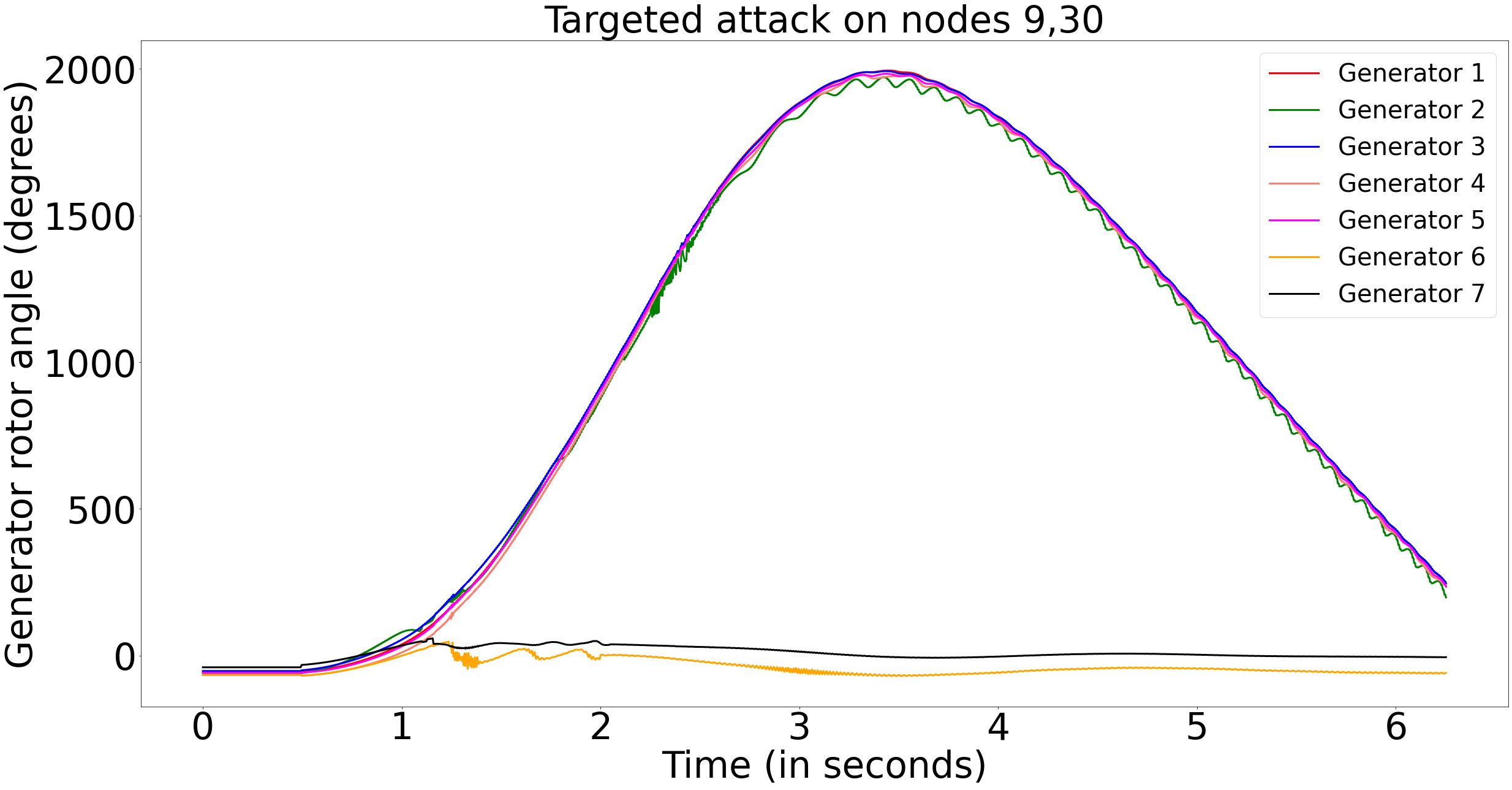}
	\caption{Plots showing the rotor angle oscillation of important generators around the Washington DC region. For the \emph{Type 1} attack (left), the rotor angle oscillation stabilizes quickly. For a \emph{Type 2} attack (right), the oscillations are severe and some generators are \emph{out-of-step} with the other generators.}
	\label{fig:geog-tar-angle-comp}
\end{figure}

\noindent\textbf{Consequences of active power deficit.} The active power generation is limited by the governor control in most generators. The lack of load-generation balance in the overloaded generators causes rotor angle instability. This is because overloaded generators decelerate and they operate \emph{out-of-step} with other generators (rotor angles difference exceeds 180 degrees). The out-of-step protection relays causes generator outages during such an occurrence. This, in turn, creates further load-generation imbalance due to reduced generation and thus eventually leads to system collapse. We show this comparison in Fig.~\ref{fig:geog-tar-angle-comp}. The \emph{Type 1} attack on Washington DC does not result in any sustained oscillations of the rotor angles of major generators in the area. On the contrary, the generator rotor angles go \emph{out-of-step} in the case of the \emph{Type 2} targeted attack.

\begin{figure}[tbhp]
    \centering
	\includegraphics[width=0.46\textwidth]{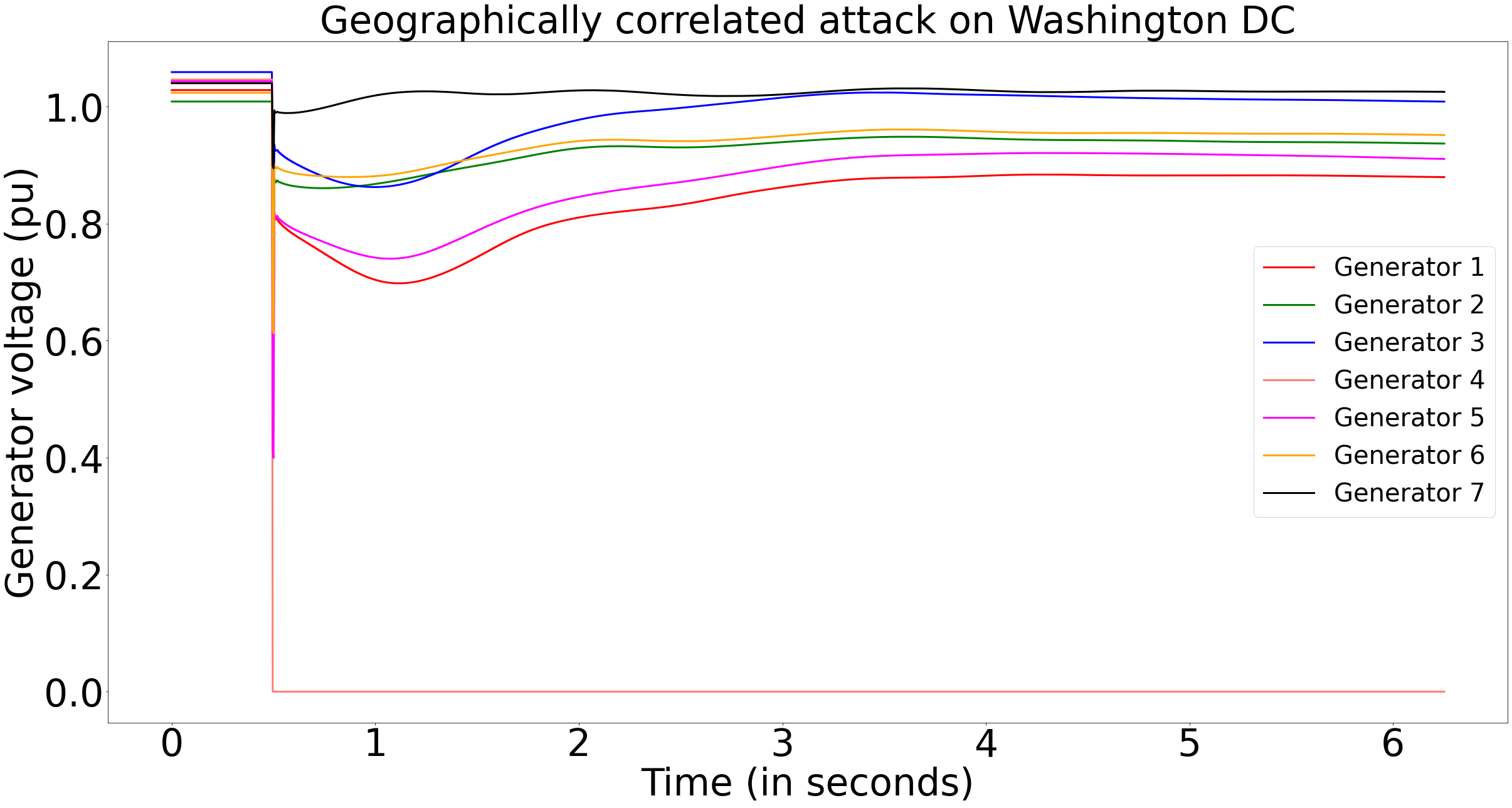}
	\includegraphics[width=0.46\textwidth]{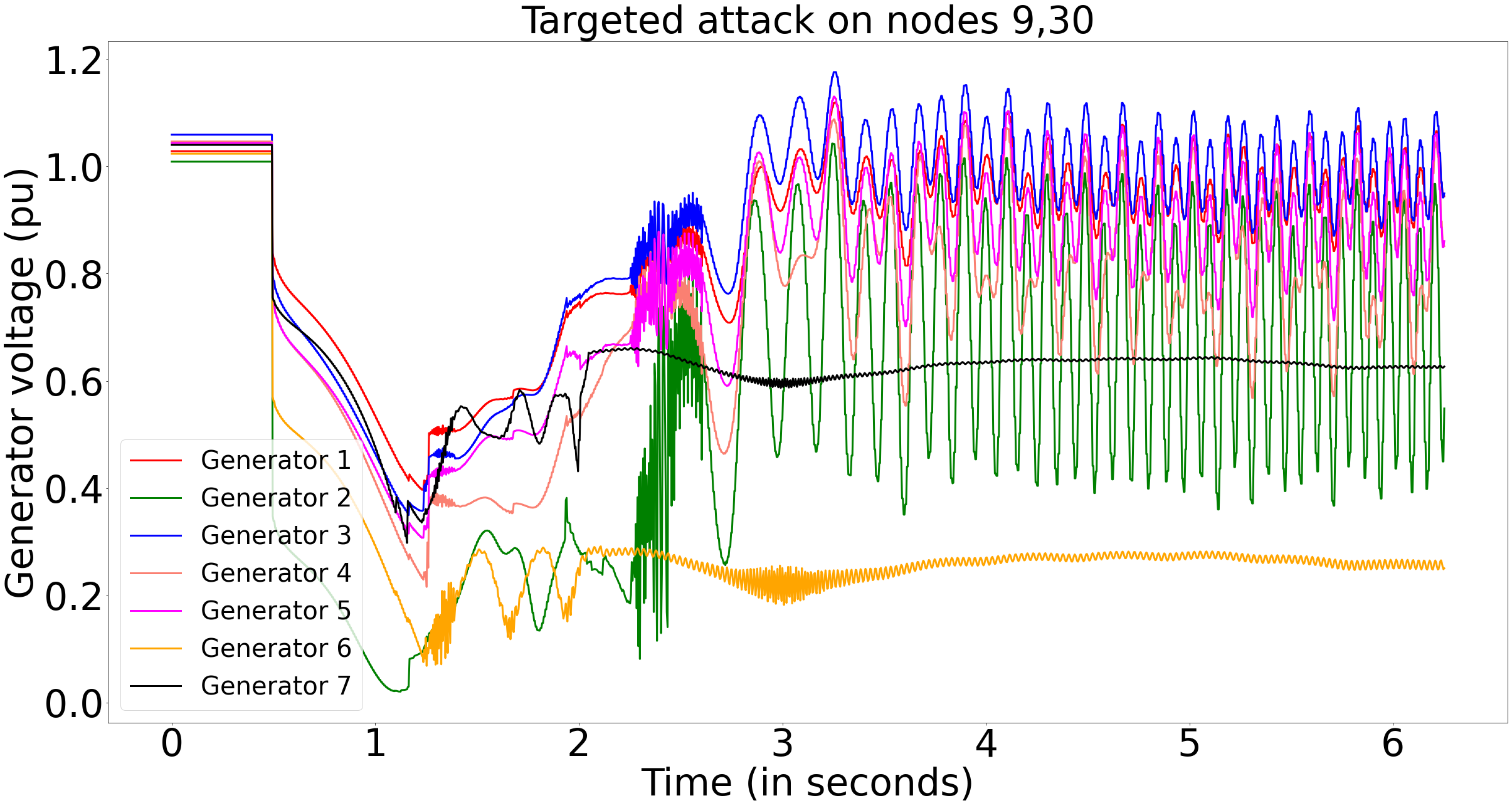}
	\caption{Plots showing the voltage oscillation of important generators near the Washington DC region. For the \emph{Type 1} attack (left), the voltage oscillation stabilizes. For a \emph{Type 2} attack (right), we observe voltage instability issues.}
	\label{fig:geog-tar-volt-comp}
\end{figure}

\noindent\textbf{Consequences of reactive power deficit.} The reactive power generation is limited by the field excitation of generators. The overcurrent protection in field circuit results in outages of the overexcited generators. The lack of reactive power support causes further voltage collapse, more line outages and eventually leading to voltage instability and eventual system collapse as depicted for the case of targeted attack in Fig.~\ref{fig:geog-tar-volt-comp}. In case of the \emph{Type 1} attack, the voltage oscillations stabilizes within a short duration.

\begin{observation}
Increasing the number of nodes in the target set does not increase the impact of attack. Additionally, the power flow magnitude along transmission line is not the primary determinant of the occurrence of a line outage.
\end{observation}
We observe the outage of a significant high voltage (500kV) transmission line connecting Washington DC and carrying almost 250MVA. We consider three different scenarios of \emph{Type 2} attack for the purpose of comparison. The choice of the nodes in the target set is based on criticality analysis performed in~\cite{rounak2018}. The necessary details are provided in the SI.
\begin{itemize}
    \item \emph{Scenario 1}: The target set consists of a single node (Target ID 9). It is known to produce maximum impact resulting in the occurrence of a system collapse when targeted singularly. 
    \item \emph{Scenario 2}: The target set consists of two nodes Target IDs 9,25. It has been observed to result in the minimum impact when Target ID 9 is combined with other 500kV nodes.
    \item \emph{Scenario 3}: The target set consists of two nodes Target IDs 9,30. This produces the maximum impact when Target ID 9 is combined with other 500kV nodes.
\end{itemize}
\begin{figure}[ht]
	\centering
	\includegraphics[width=0.46\textwidth]{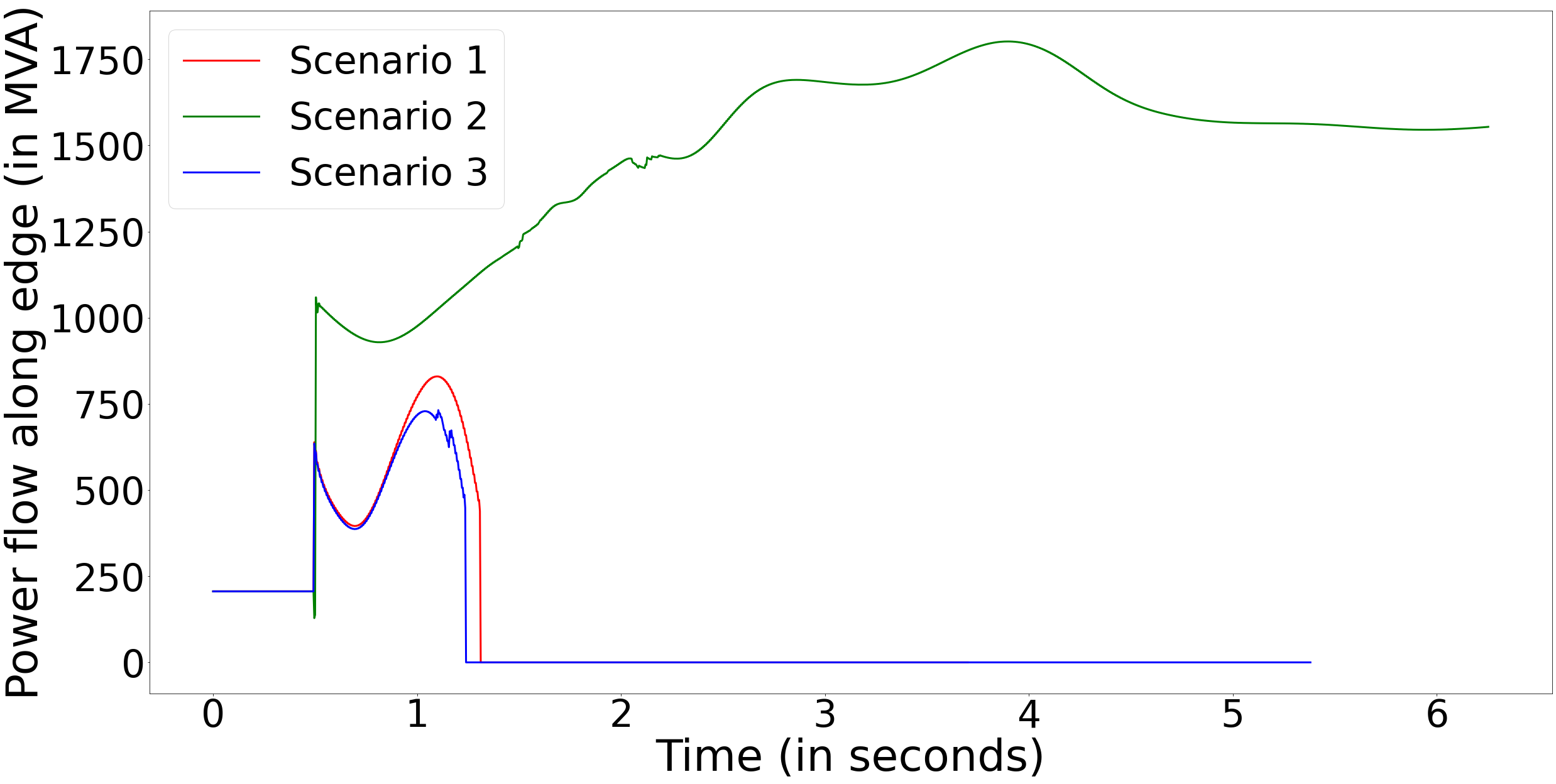}
	\includegraphics[width=0.46\textwidth]{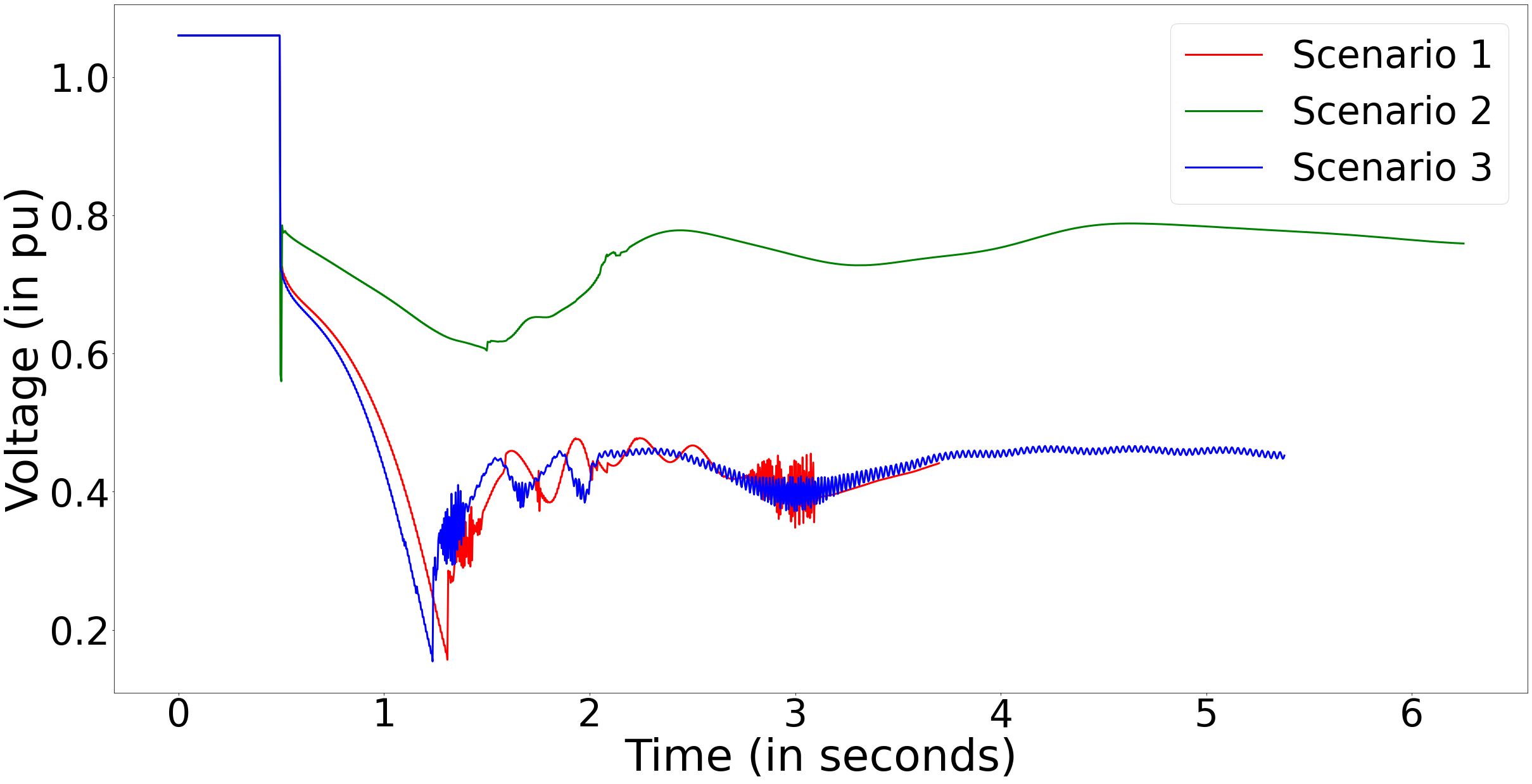}
	\includegraphics[width=0.46\textwidth]{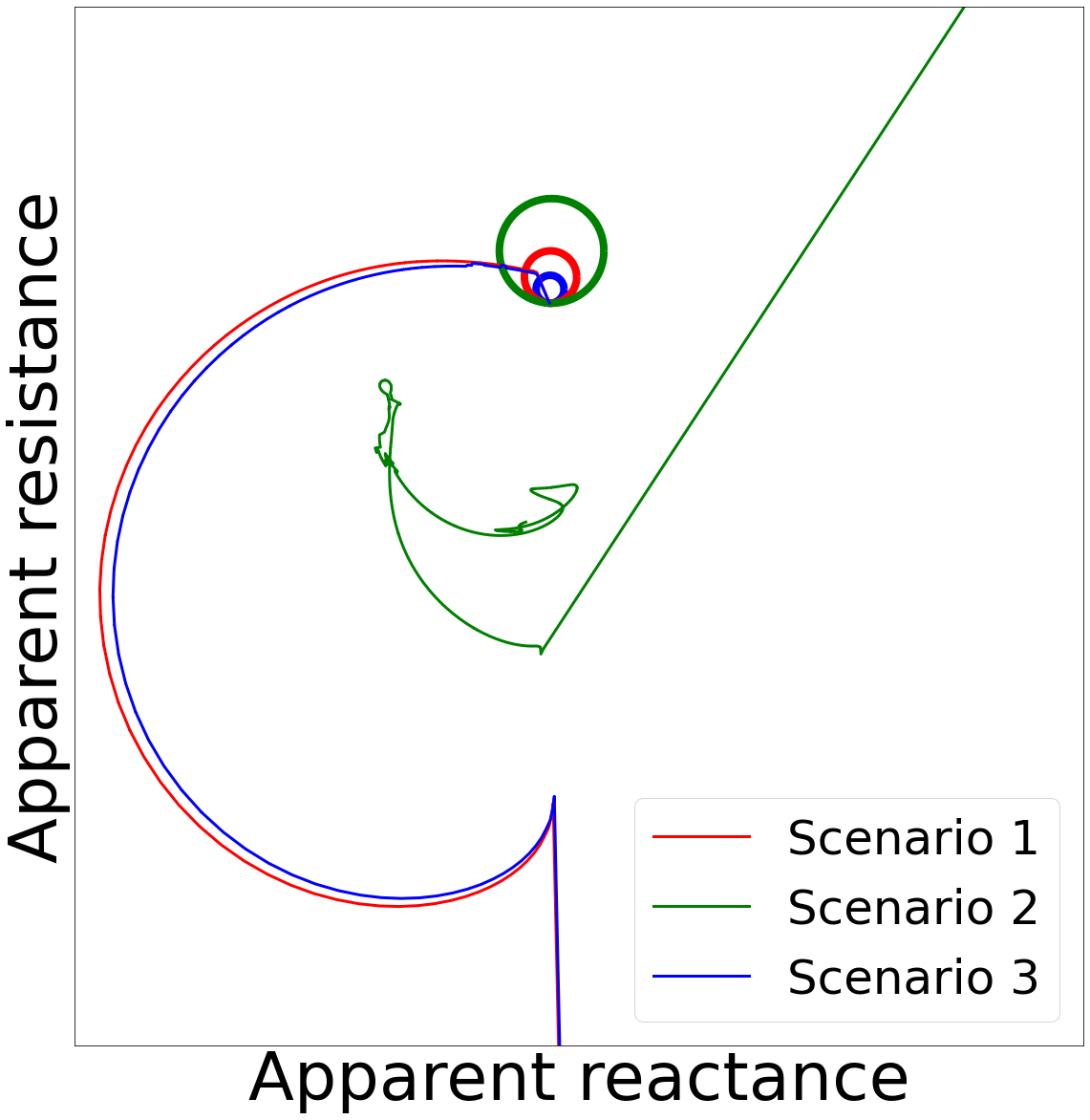}
	\includegraphics[width=0.46\textwidth]{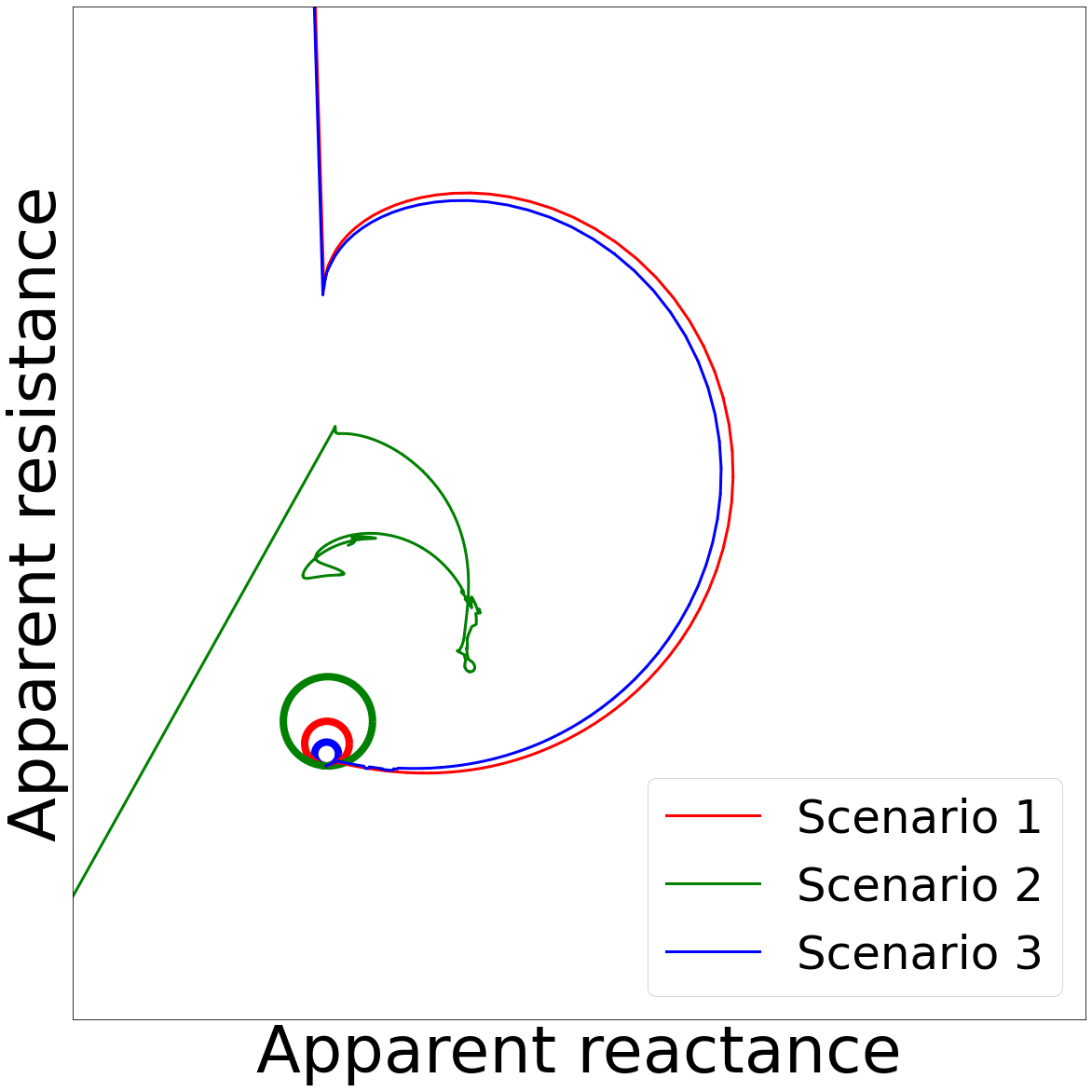}
	\caption{Plot showing variation of edge flow in a single transmission line (top left), node voltage at one of its end (top right), and apparent impedance trajectory measured by mho relays on two ends (bottom) for three targeted attack scenarios. Each color depicts a different targeted attack scenario. The node voltage drops considerably for Scenarios 1 and 3 as compared to Scenario 2. Apparent impedance trajectory enters the zone characteristics of the mho relays for Scenarios 1 and 3 which results in line trips. Scenario 3 results in the maximum power flow along the line, yet the mho relays do not trip.}
	\label{fig:line-A}
\end{figure}
The apparent power (in MVA) flowing through the transmission line in the three scenarios is shown in the left plot of Fig.~\ref{fig:line-A}. Note that after the targeted attack (at $t=1$s), the power flow along the line increases in all the three scenarios. However, the line outage occurs in Scenarios 1 and 3 (at $t=2.5$s), but there is no outage in Scenario 2. Further, we note that among the three scenarios considered in this example, the power flow along the transmission line is maximum for Scenario 2; yet it does not result in a line outage.

The apparent impedance trajectories as measured by a \emph{mho relay} installed to protect the transmission line is shown for the three scenarios in the middle plot of Fig.~\ref{fig:line-A}. While the apparent impedance encroaches the zones of protection (shown by the circular regions) for Scenarios 1 and 3, it is far away from the zones for Scenario 2. This is because of the voltage instability that occurs for Scenarios 1 and 3. We show the voltage profile at one of the connecting buses in the top-right plot of Fig.~\ref{fig:line-A}. We note that the voltage drops to a significantly low value which resulted in the apparent impedance to encroach the zones of protection.

In typical power grids, operation of transmission line relays is not dependent on power flow measured at one of its ends. It is rather protected using mho relays, directional overcurrent relays and carrier based directional comparison block relays. The operation of these relays are principal determinant of transmission line outages and are instrumental in causing cascading outages. Majority of prior works~\cite{carreras01,carreras02,carreras03,carreras04,chen_thorpe_dobson} model cascading outages using probabilities assigned based on power flowing through lines. Such approximate models are often reasonable due to dependence of current and power flowing through a line. However, they fail to represent the non-linear relationships between the power engineering quantities (current, voltage, power and impedance). A realistic representation of relay operation is necessary in the cascading failure models in order to capture the complex dynamics of cascading outages. Being devoid of such models, prior works often fail to accurately identify possible cascading outages. For instance, in the above example, a power flow based line outage model would result in a line trip for Scenario 2, since the post-event power flow through the line is maximum among the three scenarios. However, our proposed framework with realistic representation of relays show that the line outage occurs for the other scenarios, and not Scenario 2.

\begin{figure}[tbhp]
    \centering
	\includegraphics[width=0.46\textwidth]{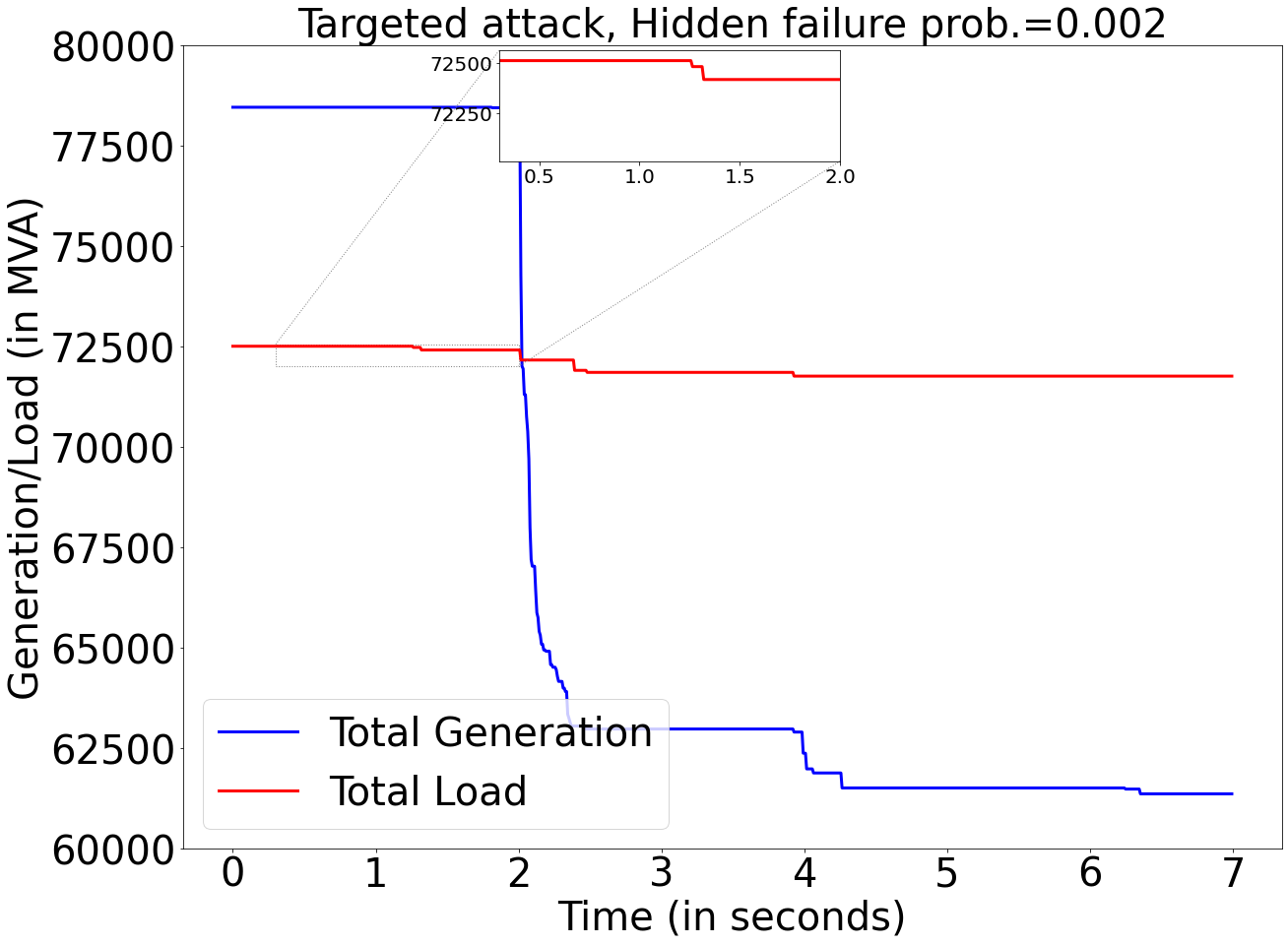}
	\includegraphics[width=0.46\textwidth]{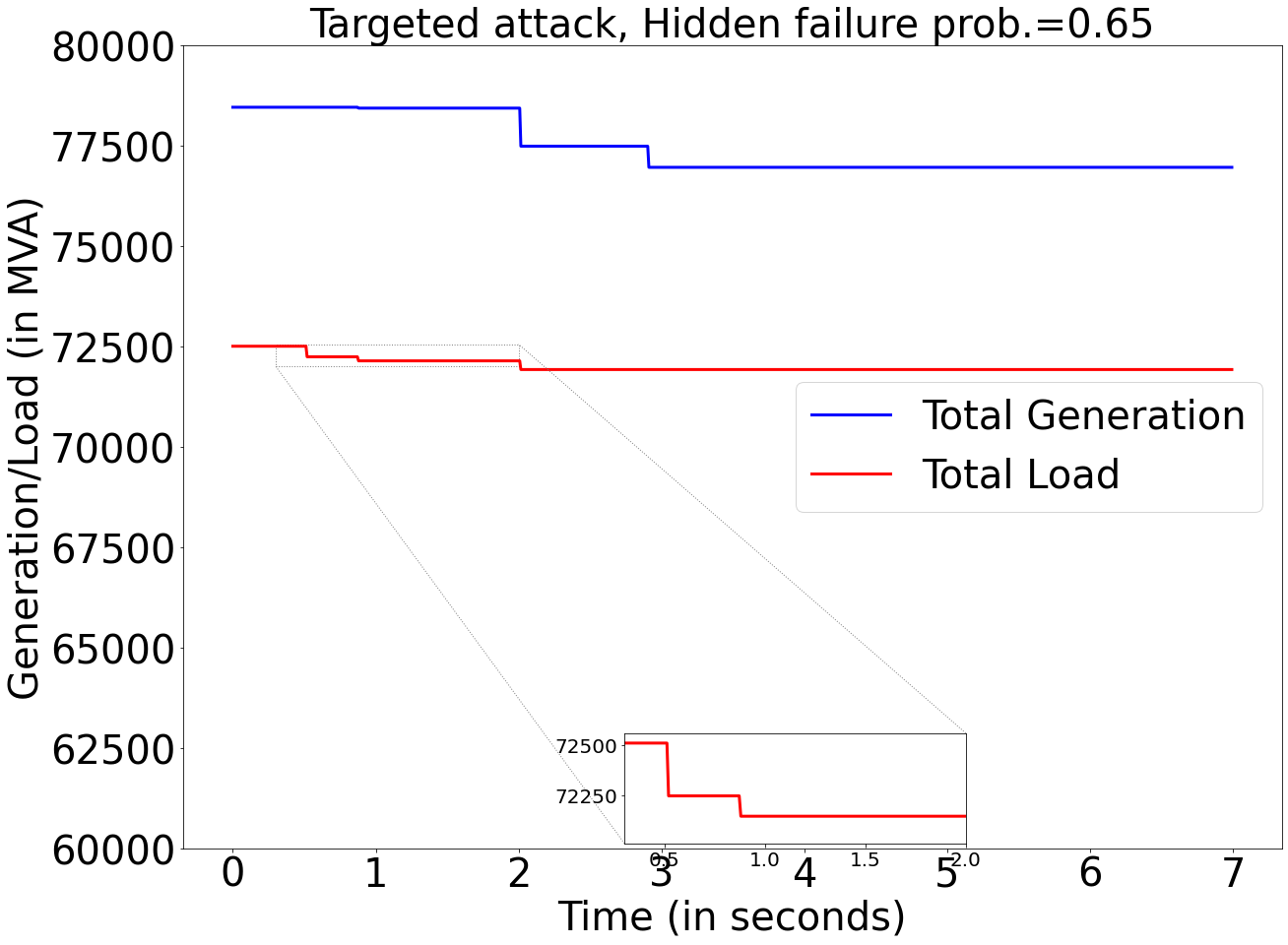}
	\caption{Plots showing change in total system load and generation in the aftermath of a targeted attack on a strategically selected substation node with low (left) and high (right) probability of hidden failure occurrence in relays. For a high probability of hidden failure, a number of transmission line outages occur following the targeted attack which causes a large amount of load getting isolated in the network, which eventually facilitates in maintaining load-generation balance.}
	\label{fig:tar-comp-hf}
\end{figure}

\begin{observation}
With an increase in occurrence of hidden failures within protection system, the extent of cascading outages reduces.
\end{observation}
To elaborate this observation, we consider Scenario 3 where simultaneous targeted attack is performed on Target IDs 9,30. We study the impact of two different probabilities ($0.2\%$ and $65\%$) of hidden failure occurrence in transmission line protection relays. Figure~\ref{fig:tar-comp-hf} compares the evolution of net load and generation in the region for the two cases as the cascading outages propagate over few seconds after the attack. The inset figures show the drop in load immediately after the physical attack occurs. A high probability of hidden failure occurrence results in the outage of a number of transmission lines immediately after the attack, leading to disconnection of loads from the power grid. This does not happen for low hidden failure occurrence probability. The disconnection of significant amount of load helps in maintaining the load-generation balance of the disturbed power grid. Therefore, the operating generators do not suffer from rotor angle and voltage instabilities.

A $65\%$ probability of occurrence of hidden failures is an extremely pessimistic estimate even for the worst maintained power grids, and hence the premise of our analysis might be an unlikely to happen in practice. Yet, our results show an important observation that an immediate disconnection of large loads from the power grid or operating major load centers as self sustaining microgrid networks can avoid a system wide collapse and a possible widespread blackout. In our analysis, the immediate disconnection of large loads occur as a result of relay misoperations due to hidden failures. In practice these can be done by power system operators in real time, even without the occurrence of hidden failures. This necessitates development of communication aided protection systems so that the relays exchange signals among each other and preferably update protection strategies~\cite{simon2020}. Wide area measurement systems (WAMS) and wide area protection systems (WAPS) which use phasor measurement units (PMUs) will play a key role in such communication aided protection strategies. Few examples of such strategies include majority voting scheme based relay decisions and system protection schemes (SPS) which are specific to a particular system~\cite{phadke,Phadke-Horowitz}.

\subsection*{Comparison of AC and DC Cascading Models}
\begin{figure}[tbhp]
	\centering
	\includegraphics[width=0.9\textwidth]{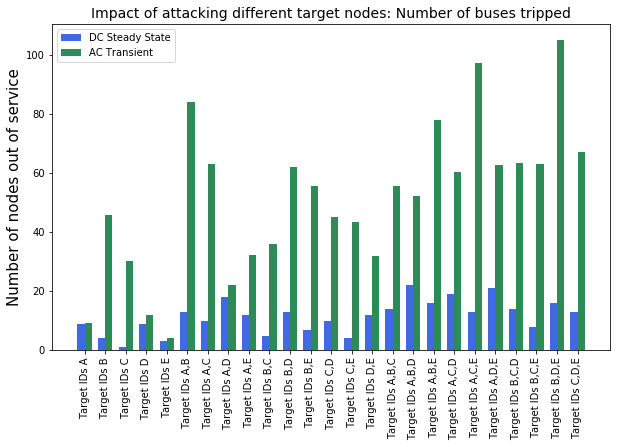}
	\caption[Impact of targeted attack on high degree nodes studied using DC steady-state analysis]{We observe the impact of targeted attack on high degree nodes using DC steady-state analysis. DC power flow based steady-state analysis underestimates the actual impact of a physical attack.}
	\label{fig:dc-imp}
\end{figure}

In this section, the traditional DC power flow based steady-state analysis is compared with the proposed cascading failure model. For the DC power flow analysis, the admittance matrix and power injections at each bus in the power system are evaluated. With the usual assumption of flat voltage profile at each bus and neglecting reactive power, the bus voltage angles are estimated. A transmission line is tripped if the electrical angular separation between the ends is more than $70$ degrees as used in~\cite{NERC_TPL}. The same set of process is executed until there are no more outages. A node with no edges connected to it is considered as a tripped node. Fig.~\ref{fig:dc-imp} depicts the impact of targeted attack on high degree nodes in the network. In comparison to the proposed model, the DC steady-state analysis underestimates the number of node outages. Furthermore, it does not identify a power system collapse due to transient instability.

\section*{Conclusions}
We built a framework to assess vulnerability of the power grid to cascading outages when subjected to a severe disturbance. The framework includes a detailed realistic representation of generation, load and transmission lines of the power grid and uses AC power flow based transient stability analysis to study the cascading outages. The operation of protection relays is considered as the principal determinant of cascading outages. Therefore, we use detailed models of traditional electro-mechanical/digital relays, and consider occurrences of probable hidden failures in them. Our generic framework allows us to perform vulnerability analysis of a given power grid initiated by any severe event, such as hurricane, earthquake, forest fires or a targeted cyber/physical attack. We show the necessity of realistic representation of protection systems to capture power grid dynamics as accurately as possible.

Particularly, we have analyzed the impact of a targeted adversarial physical attack on the power grid of Washington DC. We compare a large scale physical attack (\emph{Type 1} attack) initiated due to detonation of a tactical device at a large geographic region with a strategically targeted simultaneous attack on selected substations located far apart (\emph{Type 2} attack). Our results show that albeit severe societal damage encompassing a large region, a \emph{Type 1} attack on Washington DC area does not result in cascading power outages. 
On the contrary, a strategic simultaneous attack on selected substations can result in widespread cascading outages in and around Washington DC region. Though physically less severe than the former, this attack can impact a larger population through the resulting cascading power outage.

\clearpage
\bibliographystyle{unsrt}
\bibliography{ref_books,ref_papers,ref_reports}

\begin{thebibliography}{10}

\bibitem{clinton}
{The White House}.
\newblock {W}hite {P}aper: The {C}linton administration's policy on critical
  infrastructure protection: Presidential decision directive.
\newblock
  \url{https://clintonwhitehouse4.archives.gov/WH/EOP/NSC/html/documents/NSCDoc3.html},
  1998.

\bibitem{gop}
{Republican Policy Committee}.
\newblock Infrastructure cybersecurity: The {U.S.} electric grid.
\newblock
  \url{https://www.rpc.senate.gov/policy-papers/infrastructure-cybersecurity-the-us-electric-grid},
  July 2021.

\bibitem{nerc2021}
{North American Electric Reliability Corporation}.
\newblock State of reliability: An assessment of 2020 bulk power system
  performance.
\newblock
  \url{https://www.nerc.com/pa/RAPA/PA/Performance\%20Analysis\%20DL/NERC_SOR_2021.pdf},
  Aug 2021.

\bibitem{NERC_2016}
{Protection System Misoperations Task Force}.
\newblock {State of Reliability 2016}, May 2016.

\bibitem{NERC_2015}
{Protection System Misoperations Task Force}.
\newblock {Analysis of System Protection Misoperations}, Dec 2015.

\bibitem{NERC_2014}
{Protection System Misoperations Task Force}.
\newblock {NERC Staff Analysis of System Protection Misoperations}, Dec 2014.

\bibitem{NERC_2013}
{Protection System Misoperations Task Force}.
\newblock {Misoperations Report}, Apr 2013.

\bibitem{onl_2019}
M.~Dumas, B.~Kc, and C.~I. Cunliff.
\newblock Extreme weather and climate vulnerabilities of the electric grid: A
  summary of environmental sensitivity quantification methods.
\newblock \url{https://info.ornl.gov/sites/publications/Files/Pub128663.pdf},
  Aug 2019.

\bibitem{Weiss2019}
M.~Weiss and M.~Weiss.
\newblock An assessment of threats to the {A}merican power grid.
\newblock {\em Energy, Sustainability and Society}, 9(1):18, May 2019.

\bibitem{amin2020}
B.M. Amin, S.~Taghizadeh, Md.~S. Rahman, Md.~J. Hossain, V.~Varadharajan, and
  Z.~Chen.
\newblock Cyber attacks in smart grid – dynamic impacts, analyses and
  recommendations.
\newblock {\em IET Cyber-Physical Systems: Theory \& Applications},
  5(4):321--329, 2020.

\bibitem{liu2021}
Y.~Liu, N.~Zhang, D.~Wu, A.~Botterud, R.~Yao, and C.~Kang.
\newblock Searching for critical power system cascading failures with graph
  convolutional network.
\newblock {\em IEEE Transactions on Control of Network Systems},
  8(3):1304--1313, 2021.

\bibitem{low2021}
L.~Guo, C.~Liang, A.~Zocca, S.~H. Low, and A.~Wierman.
\newblock Line failure localization of power networks part i: Non-cut outages.
\newblock {\em IEEE Transactions on Power Systems}, 36(5):4140--4151, 2021.

\bibitem{Schafer2018}
B.~Sch\"{a}fer, D.~Witthaut, M.~Timme, and V.~Latora.
\newblock Dynamically induced cascading failures in power grids.
\newblock {\em Nature Communications}, 9(1):1975, May 2018.

\bibitem{rounak2018}
R.~Meyur, A.~Vullikanti, M.~V. Marathe, A.~Pal, M.~Youssef, and V.~Centeno.
\newblock Cascading effects of targeted attacks on the power grid.
\newblock In {\em Complex Networks and Their Applications VII}, pages 155--167,
  Dec 2018.

\bibitem{barrett_2012}
C.~Barrett, R.~Beckman, K.~Channakeshava, F.~Huang, V.~S.~A. Kumar, A.~Marathe,
  M.~V. Marathe, G.~Pei, and S.~Saha.
\newblock Human initiated cascading failures in societal infrastructures.
\newblock {\em Public Library of Science}, 7(10):1--20, Oct 2012.

\bibitem{setola_2008}
S.~Panzieri and R.~Setola.
\newblock Failure propagation in critical interdependent infrastructures.
\newblock {\em International Journal of Modeling, Identification and Control},
  3(1):69--78, May 2008.

\bibitem{texas2021}
C.W. King, J.D. Rhodes, and J.~Zarnikau.
\newblock The timeline and events of the {F}ebruary 2021 {T}exas electric grid
  blackouts.
\newblock
  \url{https://energy.utexas.edu/sites/default/files/UTAustin\%20\%282021\%29\%20EventsFebruary2021TexasBlackout\%2020210714.pdf},
  Jul 2021.

\bibitem{busby2021}
J.~W. Busby, K.~Baker, M.~D. Bazilian, A.~Q. Gilbert, E.~Grubert, V.~Rai, J.~D.
  Rhodes, S.~Shidore, C.~A. Smith, and M.~E. Webber.
\newblock Cascading risks: Understanding the 2021 winter blackout in {T}exas.
\newblock {\em Energy Research \& Social Science}, 77:102106, 2021.

\bibitem{victoria2021}
J.~Feuerstein.
\newblock Thousands of {V}ictorian homes still without power a week after winds
  and storm.
\newblock
  \url{https://7news.com.au/weather/melbourne-weather/thousands-of-victorian-homes-still-without-power-a-week-after-winds-and-storm-c-4426169},
  Nov 2021.

\bibitem{canada2022}
J.~Feuerstein.
\newblock Deadly thunderstorm complex cuts power to nearly a million in
  {C}anada.
\newblock
  \url{https://www.washingtonpost.com/weather/2022/05/22/canada-storm-derecho-ontario-quebec/},
  May 2022.

\bibitem{India}
J.~J. Romero.
\newblock Blackouts illuminate {I}ndia's power problems.
\newblock {\em IEEE Spectrum}, 49(10):11--12, Oct 2012.

\bibitem{kundur}
P.~Pourbeik, P.~S. Kundur, and C.~W. Taylor.
\newblock The anatomy of a power grid blackout - root causes and dynamics of
  recent major blackouts.
\newblock {\em IEEE Power and Energy Magazine}, 4(5):22--29, Sept 2006.

\bibitem{2003_bout}
{US-Canada Power System Outage Task Force}.
\newblock Final report on the august 14, 2003 blackout in the {U}nited {S}tates
  and {C}anada: Causes and recommendations, Apr 2004.

\bibitem{WSCCout}
D.~N. Kosterev, C.~W. Taylor, and W.~A. Mittelstadt.
\newblock {Model validation for the August 10, 1996 WSCC system outage}.
\newblock {\em IEEE Transactions on Power Systems}, 14(3):967--979, Aug 1999.

\bibitem{carreras01}
B.~A. Carreras, D.~E. Newman, I.~Dobson, and A.~B. Poole.
\newblock Initial evidence for self-organized criticality in electric power
  system blackouts.
\newblock In {\em Proceedings of the 33rd Annual Hawaii International
  Conference on System Sciences}, Jan 2000.

\bibitem{carreras02}
I.~Dobson, B.~A. Carreras, V.~E. Lynch, and D.~E. Newman.
\newblock An initial model for complex dynamics in electric power system
  blackouts.
\newblock In {\em Proceedings of the 34th Annual Hawaii International
  Conference on System Sciences}, pages 710--718, Jan 2001.

\bibitem{carreras03}
B.~A. Carreras, V.~E. Lynch, I.~Dobson, and D.~E. Newman.
\newblock Dynamics, criticality and self-organization in a model for blackouts
  in power transmission systems.
\newblock In {\em Proceedings of the 35th Annual Hawaii International
  Conference on System Sciences}, Jan 2002.

\bibitem{carreras04}
B.~A. Carreras, V.~E. Lynch, I.~Dobson, and D.~E. Newman.
\newblock Critical points and transitions in an electric power transmission
  model for cascading failure blackouts.
\newblock {\em Chaos: An Interdisciplinary Journal of Nonlinear Science},
  12(4):985--994, Sept 2002.

\bibitem{dobson2}
I.~Dobson, J.~Chen, J.~S. Thorp, B.~A. Carreras, and D.~E. Newman.
\newblock Examining criticality of blackouts in power system models with
  cascading events.
\newblock In {\em Proceedings of the 35th Annual Hawaii International
  Conference on System Sciences}, Jan 2002.

\bibitem{chen_thorpe_dobson}
J.~Chen, J.~S. Thorp, and I.~Dobson.
\newblock Cascading dynamics and mitigation assessment in power system
  disturbances via a hidden failure model.
\newblock {\em International Journal of Electrical Power \& Energy Systems},
  27(4):318 -- 326, May 2005.

\bibitem{zhang_2016}
Y.~Zhang and O.~Yağan.
\newblock Optimizing the robustness of electrical power systems against
  cascading failures.
\newblock {\em Scientific Reports}, 6:27635 1--15, Jun 2016.

\bibitem{pahwa}
S.~Pahwa, C.~Scoglio, and A.~Scala.
\newblock Abruptness of cascade failures in power grids.
\newblock {\em Scientific Reports}, 4(1):3694 1--9, Jan 2014.

\bibitem{soltan_2014}
S.~Soltan, D.~Mazauric, and G.~Zussman.
\newblock Cascading failures in power grids: Analysis and algorithms.
\newblock In {\em {Proceedings of the 5th International Conference on Future
  Energy Systems}}, e-Energy '14, pages 195--206, New York, NY, USA, Jun 2014.
  ACM.

\bibitem{Bernstein_2012}
A.~Bernstein, D.~Bienstock, D.~Hay, M.~Uzunoglu, and G.~Zussman.
\newblock Sensitivity analysis of the power grid vulnerability to large-scale
  cascading failures.
\newblock {\em SIGMETRICS Performance Evaluation Review}, 40(3):33--37, Jan
  2012.

\bibitem{stability2021}
Nikos Hatziargyriou, Jovica Milanovic, Claudia Rahmann, Venkataramana Ajjarapu,
  Claudio Canizares, Istvan Erlich, David Hill, Ian Hiskens, Innocent Kamwa,
  Bikash Pal, Pouyan Pourbeik, Juan Sanchez-Gasca, Aleksandar Stankovic,
  Thierry Van~Cutsem, Vijay Vittal, and Costas Vournas.
\newblock Definition and classification of power system stability – revisited
  \& extended.
\newblock {\em IEEE Transactions on Power Systems}, 36(4):3271--3281, 2021.

\bibitem{hines_2010}
P.~Hines, E.~Cotilla-Sanchez, and S.~Blumsack.
\newblock Do topological models provide good information about electricity
  infrastructure vulnerability?
\newblock {\em Chaos: An Interdisciplinary Journal of Nonlinear Science},
  20(3):033122 1--5, Sept 2010.

\bibitem{hines_2010_other}
P.~Hines, S.~Blumsack, E.~Cotilla Sanchez, and C.~Barrows.
\newblock The topological and electrical structure of power grids.
\newblock In {\em 2010 43rd Hawaii International Conference on System
  Sciences}, pages 1--10, 2010.

\bibitem{hines_2016}
J.~Song, E.~Cotilla-Sanchez, G.~Ghanavati, and P.~Hines.
\newblock Dynamic modeling of cascading failure in power systems.
\newblock {\em IEEE Transactions on Power Systems}, 31(3):2085--2095, May 2016.

\bibitem{stanley_2010}
S.~V. Buldyrev, R.~Parshani, G.~Paul, H.~E. Stanley, and S.~Havlin.
\newblock Catastrophic cascade of failures in interdependent networks.
\newblock {\em Nature}, 464(1):1025--1028, Apr 2010.

\bibitem{roni_2010}
Roni Parshani, Sergey~V. Buldyrev, and Shlomo Havlin.
\newblock Interdependent networks: Reducing the coupling strength leads to a
  change from a first to second order percolation transition.
\newblock {\em Physical Review Letters}, 105(4):048701 1--4, Jul 2010.

\bibitem{stanley_2011}
X.~Huang, J.~Gao, S.~V. Buldyrev, S.~Havlin, and H.~E. Stanley.
\newblock Robustness of interdependent networks under targeted attack.
\newblock {\em Physical Review E}, 83(6):065101 1--4, Jun 2011.

\bibitem{raissa_2012}
C.~D. Brummitt, R.~M. D{\textquoteright}Souza, and E.~A. Leicht.
\newblock Suppressing cascades of load in interdependent networks.
\newblock {\em Proceedings of the National Academy of Sciences},
  109(12):E680--E689, Mar 2012.

\bibitem{stanley_2018}
W.~Wang, S.~Yang, F.~Hu, H.~E. Stanley, S.~He, and M.~Shi.
\newblock An approach for cascading effects within critical infrastructure
  systems.
\newblock {\em Physica A: Statistical Mechanics and its Applications},
  510(1):164 -- 177, Nov 2018.

\bibitem{hill2020}
Z.~Wang, G.~Chen, L.~Liu, and D.~J. Hill.
\newblock Cascading risk assessment in power-communication interdependent
  networks.
\newblock {\em Physica A: Statistical Mechanics and its Applications},
  540:120496, 2020.

\bibitem{Son_2012}
Seung-Woo Son, Golnoosh Bizhani, Claire Christensen, Peter Grassberger, and
  Maya Paczuski.
\newblock Percolation theory on interdependent networks based on epidemic
  spreading.
\newblock {\em {EPL} (Europhysics Letters)}, 97(1):16006, jan 2012.

\bibitem{buldyrev2020}
Lucas~D Valdez, Louis Shekhtman, Cristian~E La~Rocca, Xin Zhang, Sergey~V
  Buldyrev, Paul~A Trunfio, Lidia~A Braunstein, and Shlomo Havlin.
\newblock Cascading failures in complex networks.
\newblock {\em Journal of Complex Networks}, 8(2), 05 2020.
\newblock cnaa013.

\bibitem{sandpile2}
Per Bak, Chao Tang, and Kurt Wiesenfeld.
\newblock Self-organized criticality.
\newblock {\em Phys. Rev. A}, 38:364--374, Jul 1988.

\bibitem{tamronglak1}
S.~Tamronglak, S.~H. Horowitz, A.~G. Phadke, and J.~S. Thorp.
\newblock Anatomy of power system blackouts: Preventive relaying strategies.
\newblock {\em IEEE Transactions on Power Delivery}, 11(2):708--715, Apr 1996.

\bibitem{tamronglak2}
J.~S. Thorp, A.~G. Phadke, S.~H. Horowitz, and S.~Tamronglak.
\newblock Anatomy of power system disturbances: Importance sampling.
\newblock {\em International Journal of Electrical Power \& Energy Systems},
  20(2):147 -- 152, Feb 1998.

\bibitem{simon2020}
Edward Schrom, Ann Kinzig, Stephanie Forrest, Andrea~L. Graham, Simon~A. Levin,
  Carl~T. Bergstrom, Carlos Castillo-Chavez, James~P. Collins, Rob~J. de~Boer,
  Adam Doupé, Roya Ensafi, Stuart Feldman, Bryan T. Grenfell.~Alex Halderman,
  Silvie Huijben, Carlo Maley, Melanie Mosesr, Alan~S. Perelson, Charles
  Perrings, Joshua Plotkin, Jennifer Rexford, and Mohit Tiwari.
\newblock Challenges in cybersecurity: Lessons from biological defense systems,
  2021.

\bibitem{pskundur}
P.~S. Kundur.
\newblock {\em Power System Stability and Control}.
\newblock McGraw Hill Education, 1994.

\bibitem{sauer}
P.~W. Sauer and M.~A. Pai.
\newblock {\em Power System Dynamics and Stability}.
\newblock Prentice Hall, 1998.

\bibitem{Phadke-Horowitz}
S.~H. Horowitz and A.~G. Phadke.
\newblock {\em Power System Relaying}.
\newblock Research Studies Press, 2nd Edition, Taunton, UK, 1995.

\bibitem{park}
K.~Park.
\newblock {\em Park's Textbook of Preventive and Social Medicine, Twenty-third
  edition}.
\newblock Bhanot Publishers, 2015.

\bibitem{phadke}
A.G. Phadke and J.S. Thorpe.
\newblock {\em Synchronized Phasor Measurements and Their Applications}.
\newblock Springer, 2008.

\bibitem{NERC_TPL}
{North American Reliability Corporation}.
\newblock {NERC Reliability Standard TPL-001-1}, Apr 2012.

\end{thebibliography}

\end{document}